\documentclass[10pt,twoside,english,aps,prl,twocolumns,reprint,showpacs,preprintnumbers,superscriptaddress,nofootinbib,floatfix]{revtex4-2}
\usepackage[T1]{fontenc}
\usepackage[latin9]{inputenc}
\usepackage{color}
\usepackage{babel}
\usepackage{float}
\usepackage{bm}
\usepackage{amsmath}
\usepackage{amsthm}
\usepackage{amssymb}
\usepackage{graphicx}
\PassOptionsToPackage{normalem}{ulem}
\usepackage{ulem}
\usepackage[unicode=true,
 bookmarks=true,bookmarksnumbered=false,bookmarksopen=false,
 breaklinks=false,pdfborder={0 0 1},backref=false,colorlinks=true]
 {hyperref}
\hypersetup{pdftitle={SU4Interaction},
 pdfauthor={BNLU},
 pdfborderstyle=,allcolors=blue}

\makeatletter

\providecommand{\tabularnewline}{\\}

\usepackage{babel}
\usepackage{amsfonts}
\usepackage{latexsym}\usepackage{times}
\usepackage{slashed}

\allowdisplaybreaks[4]

\makeatother

\begin{document}

\title{Perturbative quantum Monte Carlo method for nuclear physics}
\author{Bing-Nan Lu}
\email{bnlv@gscaep.ac.cn}
\affiliation{Graduate School of China Academy of Engineering Physics, Beijing 100193,
China}
\author{Ning Li}
\affiliation{School of Physics, Sun Yat-Sen University, Guangzhou 510275, China}
\author{Serdar Elhatisari}
\affiliation{Faculty of Natural Sciences and Engineering, Gaziantep Islam Science
and Technology University, Gaziantep 27010, Turkey}
\author{Yuan-Zhuo Ma}
\affiliation{Guangdong Provincial Key Laboratory of Nuclear Science, Institute
of Quantum Matter, South China Normal University, Guangzhou 510006,
China}
\author{Dean Lee}
\email{lee.dean.j@gmail.com}
\affiliation{Facility for Rare Isotope Beams and Department of Physics and Astronomy,
Michigan State University, MI 48824, USA}
\author{Ulf-G. Mei{\ss}ner}
\email{meissner@hiskp.uni-bonn.de}
\affiliation{Helmholtz-Institut f\"ur Strahlen- und Kernphysik and Bethe Center
for Theoretical Physics, Universit\"at Bonn, D-53115 Bonn, Germany}
\affiliation{ Institute~for~Advanced~Simulation, Institut~f\"ur~Kernphysik,~
and J\"ulich~Center~for~Hadron~Physics, ~\\
 Forschungszentrum~J\"ulich, D-52425~J\"ulich,~Germany}
\affiliation{Tbilisi State University, 0186 Tbilisi, Georgia}
\begin{abstract}
While first order perturbation theory is routinely used in quantum
Monte Carlo (QMC) calculations, higher-order terms
present significant numerical challenges. We
present a new approach for computing perturbative corrections in projection
QMC calculations.
We demonstrate the method
by computing nuclear ground state energies up to second order
for a realistic chiral interaction. We calculate the binding
energies of several light nuclei up to $^{16}$O by expanding the
Hamiltonian around the Wigner SU(4) limit and find good agreement with
data. In contrast to the natural ordering of the perturbative
series, we find remarkably large second order energy corrections.
This occurs because the perturbing interactions break the symmetries
of the unperturbed Hamiltonian. Our method is free from the sign problem
and can be applied to QMC calculations for many-body systems in nuclear
physics, condensed matter physics, ultracold atoms, and quantum chemistry.
\end{abstract}
\maketitle


Quantum Monte Carlo (QMC) simulation is a powerful method
for addressing quantum many-body problems in nuclear physics \cite{Langanke:1995zz,Carlson2015_RMP,Lee2009_PPNP,Lahde:2019npb},
condensed matter \cite{Ceperley:1980zz,Foulkes2001_RMP,Assaad:2013xua},
ultracold atoms \cite{Bulgac:2008zz,Carlson:2011kv,He:2019ipt}, and
quantum chemistry \cite{Hammond1994,Nightingale1999}. Perhaps the
most important feature of QMC is that when the MC process has only positive weights,
the computational effort scales only polynomially with system size.
Unfortunately, this is not true in general.
If the Monte Carlo process
involves cancellations between positive and negative weights, the
resulting ``sign problem'' leads to exponential scaling of the computational
effort with system size. Although finding a generic solution for the
sign problem is unlikely in the near term \cite{Troyer2005_PRL},
for several important cases QMC algorithms can be applied without
sign problems, such as lattice QCD at
zero baryon density \cite{Muroya2003_PTP}, the repulsive Fermi-Hubbard
model at half-filling \cite{Varney2009_PRB}, and low-energy nuclear
systems in the Wigner SU(4) limit \cite{Wigner1937_PR,Elhatisari2016_PRL,Lu2019_PLB,Lee2021_PRL}.
The realistic systems of physical interests, though, often deviate
from these ideal models significantly and have a sign problem. In
these cases, perturbation theory can be used to bridge the difference
between the simplified  and the realistic interaction.
However, so far perturbation theory in QMC is mostly limited to the
first order. Improving the quality of the perturbative calculations
requires going to higher orders.

In Rayleigh-Schr{\"o}dinger perturbation theory, the second-order
energy correction involves a summation over all quantum states that
can be reached via the perturbing interaction. Such a calculation
over all quantum states is not compatible with QMC, which
targets only the lowest energy states. 
To solve this problem, we introduce a computational framework
called perturbative QMC (ptQMC), which allows for the efficient calculation
of higher-order perturbative corrections within the Euclidean time
formalism. As a demonstration, we implement this method using nuclear
lattice effective field theory (NLEFT) \cite{Lee2009_PPNP,Lahde:2019npb}  and perform
benchmark calculations of the binding energies of several  nuclei.


NLEFT is a QMC method for nuclear \textit{ab initio} calculations.
We regularize the chiral nuclear force on a periodic cubic lattice
and employ the auxiliary field MC method to simulate finite
nuclei. The advantage of this approach is that  many-body correlation
effects such as clustering emerge automatically \cite{Elhatisari2017_PRL,Summerfield2021_PRC}.
Due to the sign problem, early NLEFT calculations were limited to
a few 
nuclei and specially designed interactions \cite{Borasoy2007_EPJA,Borasoy2008_EPJA,Epelbaum2009_EPJA,Epelbaum2010_PRL,Epelbaum2010_EPJA,Lahde2014_PLB}.
In most of the recent NLEFT
calculations, the higher order chiral interactions are included with
first order perturbation theory \cite{Epelbaum2011_PRL,Epelbaum2012_PRL109,Epelbaum2013_PRL,Epelbaum2014_PRL,Elhatisari2015_Nature}.

The nuclear Hamiltonian is $H=K+V_{0}+V_{C}$, with $K=-\nabla^{2}/2m$
the kinetic energy operator and $m=938.92\,$MeV the nucleon mass.
We use a lattice spacing of $a=1.32$~fm.
The interaction is split into a dominant term $V_{0}$ and a correction $V_{C}$.
The ground state of $H$ can be found by applying imaginary time projectors
to a trial wave function $|\Psi_{T}\rangle$, $|\Psi\rangle=\lim_{L_{t}\to\infty}M^{L_{t}/2}|\Psi_{T}\rangle$,
with $M=:e^{-a_{t}H}:$ the transfer matrix and $a_{t}$ the temporal
step. The colons denote normal ordering. 
Without loss of generality, we assume that both $V_{0}$ and $V_{C}$
can be decomposed in terms of auxiliary fields. For example, using a simple contact
interaction for $V_{0}$,
\begin{equation}
:e^{-\tfrac{1}{2}a_{t}C_{0}\rho({\bm{n}})^{2}}: \propto\int\mathcal{D}s:e^{-\tfrac{s(\bm{n})^{2}}{2}+\sqrt{-a_{t}C_{0}}s(\bm{n})\rho(\bm{n})}:,
\end{equation}
with $\rho(\bm{n})$  the nucleon density and $s(\bm{n})$  a
real auxiliary field. We further require that $V_{0}$ does not induce
a sign problem. This is possible when $V_{0}$ is attractive with $C_{0}<0$,
and each spin-up nucleon in $|\Psi_{T}\rangle$ is paired  with
a spin-down nucleon \cite{Li2016_PRL}. This is the case for the
ground states of even-even nuclei. However, we can use a
more general $V_{C}$ that may have a sign problem. By decomposing $V_{C}$
in the same manner, we have similar expressions for the density $\rho_{c}$
and the corresponding auxiliary field $c$. For non-perturbative QMC
calculations, we need to sample both $s$ and $c$ fields.


Under the assumption that $V_{C}$ is small compared to $V_{0}$,
we can expand $|\Psi\rangle$ in powers of $V_{C}$,
\begin{eqnarray}
|\Psi\rangle & = & \lim_{L_{t}\rightarrow\infty}M^{L_{t}/2}|\Psi_{T}\rangle=|\Psi_{0}\rangle+|\delta\Psi_{1}\rangle+\mathcal{O}(V_{C}^{2}),\label{eq:expansion_of_psi}\\
|\Psi_{0}\rangle & = & \lim_{L_{t}\rightarrow\infty}M_{0}^{L_{t}/2}|\Psi_{T}\rangle,\\
|\delta\Psi_{1}\rangle & = & \lim_{L_{t}\rightarrow\infty}\sum_{k=1}^{L_{t}/2}M_{0}^{L_{t}/2-k}(M-M_{0})M_{0}^{k-1}|\Psi_{T}\rangle,
\end{eqnarray}
where $M_{0}=:e^{-a_{t}(K+V_{0})}:$ is the zeroth order transfer
matrix and we have omitted the $\mathcal{O}(a_{t}^{2})$ terms. In
Eq.~(\ref{eq:expansion_of_psi}) and what follows, we use the subscripts
to denote the perturbative orders and the symbols with $\delta$ to
represent the corrections.
The normalized wave function is
\begin{align}
|\Psi^{\prime}\rangle & =\frac{|\Psi\rangle}{\sqrt{\langle\Psi|\Psi\rangle}}=\frac{|\Psi_{0}\rangle}{\sqrt{\langle\Psi_{0}|\Psi_{0}\rangle}}+\frac{1}{\sqrt{\langle\Psi_{0}|\Psi_{0}\rangle}}\nonumber \\
 & \times\left[|\delta\Psi_{1}\rangle-\frac{{\rm Re}\langle\Psi_{0}|\delta\Psi_{1}\rangle}{\langle\Psi_{0}|\Psi_{0}\rangle}|\Psi_{0}\rangle\right]+\mathcal{O}(V_{C}^{2}),\label{eq:norm_wave-1-1}
\end{align}
where ${\rm Re}$ denotes the real part. Eq.~(\ref{eq:norm_wave-1-1})
can be used to calculate the expectation value of any operator up
to $\mathcal{O}(V_{C})$. A special case is the energy, for which
$\delta E_{1}$ only depends on $|\Psi_{0}\rangle$. With $|\delta\Psi_{1}\rangle$
at hand, we can continue further to find $\delta E_{2}$. 
The partial energy contributions at each order are
\begin{eqnarray}
E_{0} & = & \langle\Psi_{0}|(K+V_{0})|\Psi_{0}\rangle/\langle\Psi_{0}|\Psi_{0}\rangle,\\
\delta E_{1} & = & \langle\Psi_{0}|V_{C}|\Psi_{0}\rangle/\langle\Psi_{0}|\Psi_{0}\rangle,\\
\delta E_{2} & = & {\rm Re}(\langle\Psi_{0}|V_{C}|\delta\Psi_{1}\rangle-\delta E_{1}\langle\Psi_{0}|\delta\Psi_{1}\rangle)/\langle\Psi_{0}|\Psi_{0}\rangle,\label{eq:energy_pert_012-1-1}
\end{eqnarray}
where all matrix elements and overlaps can be expressed with the amplitudes,
\begin{eqnarray}
\mathcal{M}(O) & = & \langle\Psi_{T}|M_{0}^{L_{t}/2}OM_{0}^{L_{t}/2}|\Psi_{T}\rangle,\\
\mathcal{M}_{k}(O) & = & \langle\Psi_{T}|M_{0}^{L_{t}/2}OM_{0}^{L_{t}/2-k}MM_{0}^{k-1}|\Psi_{T}\rangle,
\end{eqnarray}
where $k=1,2,\cdots,L_{t}/2$. Here $O$ is the operator inserted
in the middle time step like $1,K+V_{0}$ or $V_{C}$. In $\mathcal{M}_{k}(O)$
the $k$-th copy of $M_{0}$ is replaced by the full transfer matrix
$M$. The transfer matrices $M_{0}$ and $M$ in these amplitudes
are computed using the auxiliary field formalism.

The energies $E_{0}$ and $\delta E_{1}$ are just the expectation
values $\langle O\rangle=\mathcal{M}(O)/\mathcal{M}(1)$ with $O=K+V_{0}$
or $V_{C}$. These can be calculated by sampling the auxiliary fields
$s$ in $M_{0}$ with standard algorithms \cite{Lee2009_PPNP,Lahde:2019npb}.
For $\delta E_{2}$ we need to evaluate 
an integral over the auxiliary field $c$ from the inserted $M$ in
$\mathcal{M}_{k}(O)$. For every sample $\{s_{1},s_{2},\cdots,s_{L_{t}}\}$
we have
\begin{equation}
\mathcal{M}_{k}(O)=\int\mathcal{D}cP(c+\bar{c})\langle\cdots O\cdots M(s_{k},c+\bar{c})\cdots\rangle_{T},\label{eq:shifted_integral}
\end{equation}
where the ellipses denote the transfer matrices $M_{0}(s_{t})$
with $t\neq k$, $\langle\rangle_{T}$  the expectation value in the
state $|\Psi_{T}\rangle$ and  $P(c)$ is the standard normal distribution.
In Eq.~(\ref{eq:shifted_integral}) we have made a variable change $c\rightarrow \bar{c} + c$
with $c$ real integral variables.
Here $\bar{c}(\bm{n})$ is a constant field
\begin{align}
\bar{c}(\bm{n}) & =\left.\frac{\partial}{\partial c(\bm{n})}\ln\langle\cdots M(s_{k},c)\cdots\rangle_{T}\right|_{c=0}\nonumber \\
 & =\sqrt{-a_{t}C}\langle\cdots:M_{0}(s_{k})\rho_{c}(\bm{n}):\cdots\rangle_{T}/\mathcal{M}(1),\label{eq:stationary_point_condition}
\end{align}
where the ellipses again represent the $M_{0}$'s, $C$ is the coupling
constant for the $V_{C}$ term. Generally, $\bar{c}$ is a complex
field, e.g., for repulsive interactions such as Coulomb we have $C>0$,
the square root in Eq.~(\ref{eq:stationary_point_condition}) introduces
an imaginary factor $i$. In this case the integrand in Eq.~(\ref{eq:shifted_integral})
contains non-zero phases that may induce a severe sign problem. The
variable change in Eq.~(\ref{eq:shifted_integral}) serves to alleviate
this problem \cite{Zhang2003_PRL}. To see this, we take the
logarithm of the integrand in Eq.~(\ref{eq:shifted_integral}), expand
the result near $c=0$ and apply Eq.~(\ref{eq:stationary_point_condition}).\sout{}
We find that the terms linear in $c$ and $\bar{c}$ which cause the
sign problem cancel exactly and the integrand can be factorized as
\begin{equation}
\mathcal{M}_{k}(O)=\mathcal{M}(s)\exp\left(\frac{\bar{c}^{2}}{2}\right)\int\mathcal{D}c\exp\left(-\frac{c^{2}}{2}+\epsilon\right),\label{eq:integrad_factorized}
\end{equation}
where we omit the summations over lattice sites, $\epsilon$ is a
residual term containing quadratic and higher powers of $c$.
Because in $\mathcal{M}_{k}(s,c)$ a common factor
$\sqrt{a_{t}}$ is attached to every $c$ variable, $\epsilon$ is
a small number of the order $\mathcal{O}(a_{t})$. For sufficiently
small $a_{t}$, Eq.~(\ref{eq:integrad_factorized}) means that the
integrand in Eq.~(\ref{eq:shifted_integral}) is a product of a normal
distribution and a slowly varying function $\exp(\epsilon)$. We can
use stochastic methods to evaluate Eq.~(\ref{eq:shifted_integral})
by sampling the $c$ field with a standard normal distribution. This
evaluation is unbiased and its uncertainty is determined by the variation
of $\exp(\epsilon)$. In practice, we found that the
variable change in Eq.~(\ref{eq:shifted_integral}) can reduce the
statistical error by one order or more, see~\cite{SM}
for a demonstration.


We benchmark the ptQMC using a realistic nuclear chiral force with
two-body and three-body interactions up to N$^2$LO~\cite{Epelbaum2009_RMP,Reinert2018_EPJA}.
The two-body contact terms and the one-pion-exchange potential (OPEP) read
\begin{align}
 & V_{{\rm 2N}}=\left[B_{1}+B_{2}(\bm{\sigma}_{1}\cdot\bm{\sigma}_{2})+C_{1}q^{2}+C_{2}q^{2}(\bm{\tau}_{1}\cdot\bm{\tau}_{2})\right.\nonumber \\
 & +C_{3}q^{2}(\bm{\sigma}_{1}\cdot\bm{\sigma}_{2})+C_{4}q^{2}(\bm{\sigma}_{1}\cdot\bm{\sigma}_{2})(\bm{\tau}_{1}\cdot\bm{\tau}_{2})\nonumber \\
 & +C_{5}\frac{i}{2}(\bm{q}\times\bm{k})\cdot(\bm{\sigma}_{1}+\bm{\sigma}_{2})+C_{6}(\bm{\sigma}_{1}\cdot\bm{q})(\bm{\sigma}_{2}\cdot\bm{q})\nonumber \\
 & \left.+C_{7}(\bm{\sigma}_{1}\cdot\bm{q})(\bm{\sigma}_{2}\cdot\bm{q})(\bm{\tau}_{1}\cdot\bm{\tau}_{2})\right]f_{{\rm 2N}}(p_{1},p_{2},p_{1}^{\prime},p_{2}^{\prime})\nonumber \\
 & -\frac{g_{A}^{2}f_{\pi}(q^{2})}{4F_{\pi}^{2}}\left[\frac{(\bm{\sigma}_{1}\cdot\bm{q})(\bm{\sigma}_{2}\cdot\bm{q})}{q^{2}+M_{\pi}^{2}}+C_{\pi}^{\prime}\bm{\sigma}_{1}\cdot\bm{\sigma}_{2}\right](\bm{\tau}_{1}\cdot\bm{\tau}_{2}),\label{eq:V2N}
\end{align}
where $\bm{\sigma}_{1,2} (\bm{\tau}_{1,2})$ are spin (isospin) matrices,
$B_{i}, C_{i}$ are low-energy constants (LECs).
$\bm{p}$ and $\bm{p}^{\prime}$ are the relative incoming and outgoing
momenta, respectively, $\bm{q}=\bm{p}-\bm{p}^{\prime}$, $\bm{k}=(\bm{p}+\bm{p}^{\prime})/2$
are momentum transfers, $\bm{p}_{i}$ and $\bm{p}_{i}^{\prime}$
are the momenta of the individual nucleons,
$g_{A}$, $F_{\pi}$, $M_{\pi}$ are
the axial-vector coupling constant, pion decay constant and pion mass,
respectively.
The additional regulators, $f_{{\rm 2N}}=\exp[-\sum_{i=1}^{2}\left(p_{i}^{6}+p_{i}^{\prime6}\right)/\Lambda^{6}]$ with $\Lambda=340$~MeV and $f_{\pi}=\exp[-(q^{2}+M_{\pi}^{2})/\Lambda_{\pi}^{2}]$ with $\Lambda_\pi=300$~MeV,
are introduced to minimize lattice artifacts.
For the OPEP we introduce a counterterm  $\sim C^\prime_\pi$ as in Ref.~\cite{Reinert2018_EPJA} to remove
the short-range singularity,
which, together with a low $\Lambda_\pi$, adapts the potential to perturbative calculations.
Note that the OPEP contains
a tensor interaction that couples different partial
waves, thus will contribute significantly to the energy at second
order. 
For the three-body force $V_{\rm{3N}}$ we adopt a simple 3N contact term with the LEC $c_E$.
The LECs $B_{i}, C_{i}, c_{E}$
are fixed from NN scattering data and the triton binding energy.
We also implement a static Coulomb
force $V_{{\rm cou}}$, see~\cite{SM} for further details of the interaction.


\begin{figure}
\begin{centering}
\includegraphics[width=0.95\columnwidth]{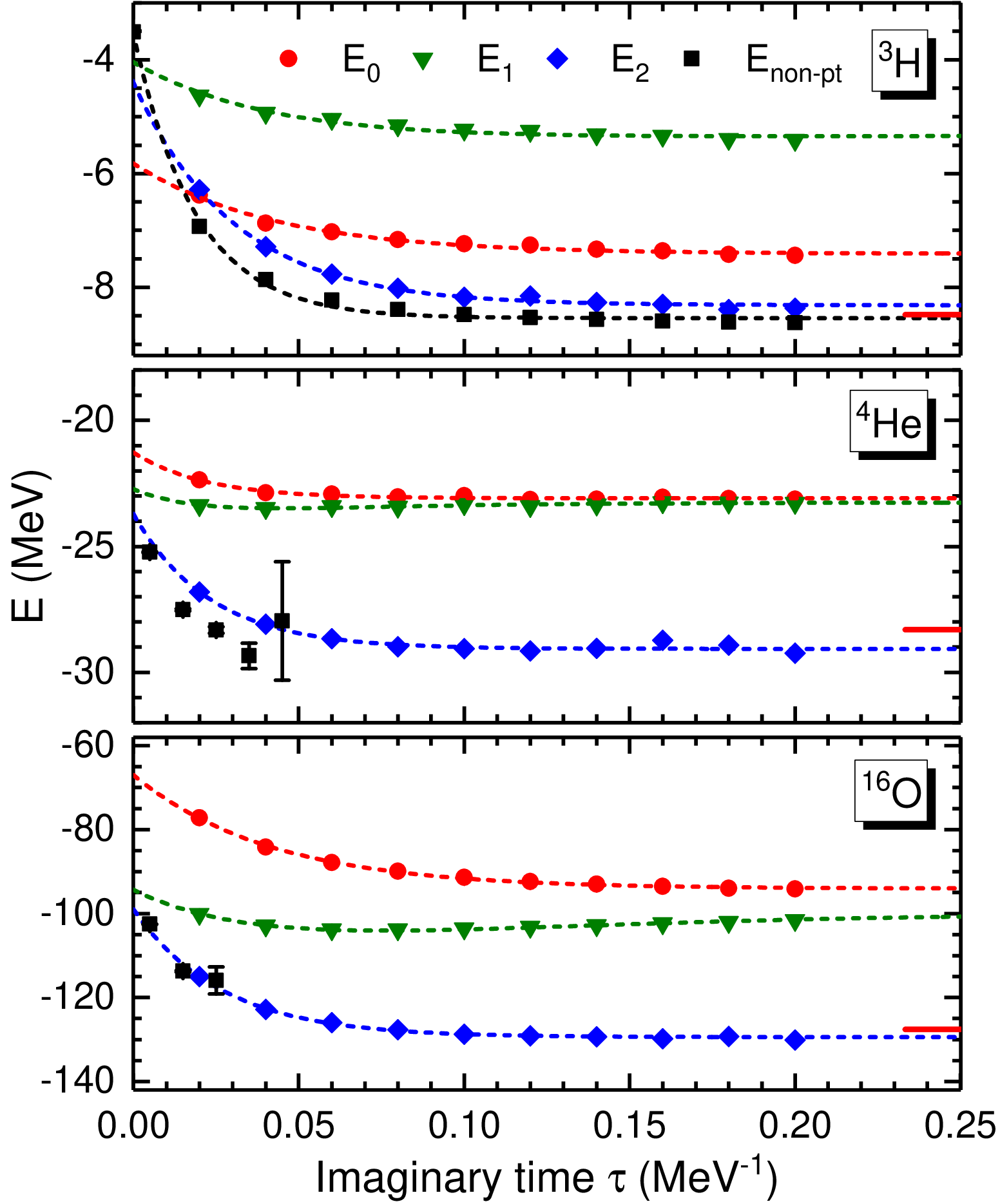}
\par\end{centering}
\caption{\label{fig:Alleviation-of-the}ptQMC binding energies as functions
of the projection time $\tau$ compared with non-perturbative results.
The circles (red), down triangles (green) and diamonds
(blue) denote the energies at the zeroth, first and second orders, respectively. The squares
(black) represent the exact results calculated with sparse matrix
multiplications for $^{3}$H and full non-perturbative QMC for $^{4}$He and $^{16}$O, respectively.
Each group of results
are fitted with a sum of exponential functions (dashed lines). The
red bars mark the experimental binding energies.}
\end{figure}

In order to compute ground states of
$H=K+V_{{\rm 2N}}+V_{{\rm 3N}}+V_{{\rm cou}}$ using ptQMC, we shall
 choose a zeroth order Hamiltonian $H_{0}=K+V_{0}$ and calculate
the energy corrections with respect to  $V_{C}=H-H_{0}$.
We take $V_{0}$ to be the non-locally smeared SU(4)
interaction from Ref.~\cite{Lu2019_PLB}, which
captures the essential elements of the nuclear
force. For benchmarking purposes, we only keep the two-body
part of $V_{0}$, which induces no sign problem for even-even nuclei.
The details of $V_{0}$ can  be found in~\cite{SM}.
For further work starting with the Wigner SU(4) limit,
see \cite{Koenig2017_PRL,Koenig2020_EPJA,Vanasse2017_FBS}.

In Fig.~\ref{fig:Alleviation-of-the} we compare the results obtained
using ptQMC with non-perturbative results. We use a periodic box of
size $L=10$ for $^{3}$H and $L=8$ for the
other nuclei. The temporal step is $a_{t}=1/1000$~MeV$^{-1}$.
For $^{3}$H, the system is small enough that we can
use exact sparse matrix calculations.
For larger nuclei 
we perform fully non-perturbative QMC calculations instead,
which result in large error bars due to severe sign problems. 
For the $^{16}$O nucleus, the sign problem sets in so quickly
that we cannot find meaningful results
to make a reliable extrapolation. However, the ptQMC calculations
are free from sign problems. The corresponding statistical errors
are smaller than the size of the symbols.
We use a sum of decaying exponential functions to capture the residual effects
of higher energy excitations and extrapolate the results to $\tau\rightarrow\infty$.
\textcolor{blue}{{}}See~\cite{SM}
for further settings of the QMC simulation.
For all three  nuclei, the second order energy corrections
are large and essential in reproducing the data. While this might
seem contrary to the normal hierarchy of the perturbative series, we
will show below that it is actually a consequence of the symmetry
breaking.

\begin{figure}
\begin{centering}
\includegraphics[width=1\columnwidth]{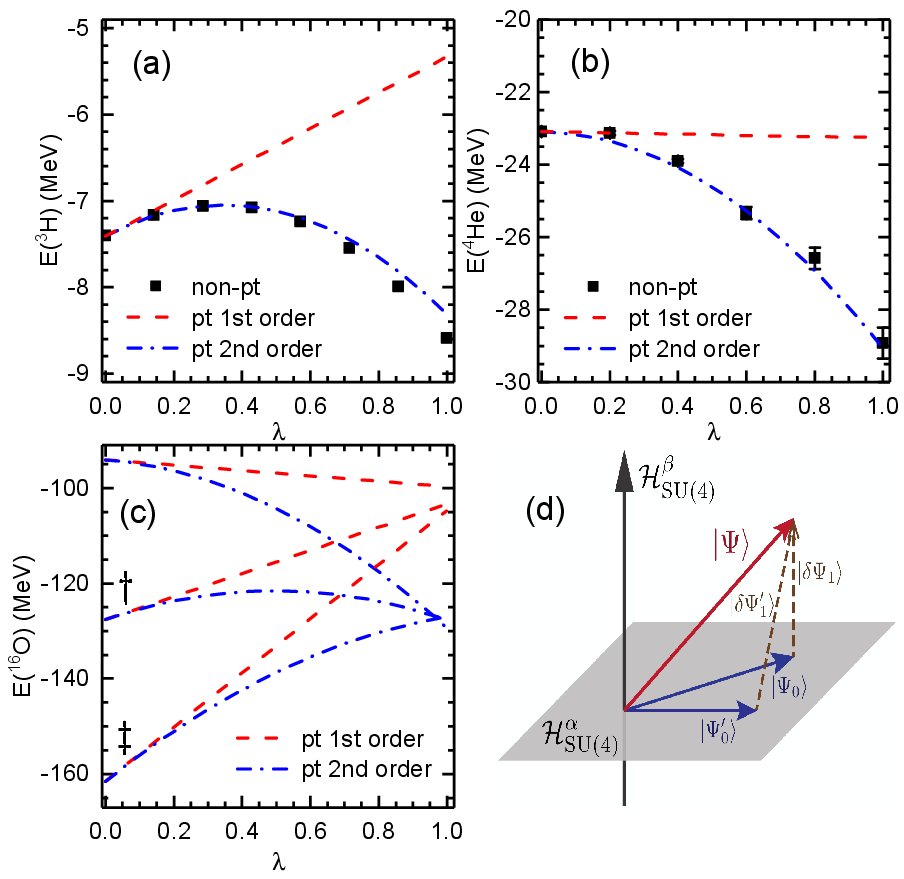}
\par\end{centering}
\caption{\label{fig:lambda_dependence}(a)(b) The dashed and dash-dotted
 lines denote the binding energies of $^{3}$H (a) and $^{4}$He (b) as functions
of the small parameter $\lambda$ in first and second order ptQMC
calculations, respectively. The black squares are the exact
results. 
(c) The first and second order ptQMC calculations for $^{16}$O, starting
from three different zeroth order interactions $V_{0}$, $1.1V_{0}$
($\dagger$) and $1.2V_{0}$ ($\ddagger$). \textcolor{blue}{}(d)
Schematic plot for a perturbative calculation. The zeroth
order wave functions $|\Psi_{0}\rangle$ and $|\Psi_{0}^{\prime}\rangle$
are confined in a subspace corresponding to an {\em irrep}
 of  SU(4).}
\end{figure}

We can now examine the convergence pattern
of the perturbative series. In Fig.~\ref{fig:lambda_dependence}(a)(b)(c)
we show the calculated energies as a function of $\lambda$, a real
number between $0$ and $1$ that we insert as a control parameter
multiplying the perturbation $V_{C}$.
The ptQMC
results are shown as lines. Because ptQMC corresponds to the Taylor
series expansion at $\lambda=0$, we find straight lines at first
and parabolas at second order.
For $^{3}$H ($^{4}$He)
we also display the exact energies of $H_{0}+\lambda V_{C}$ obtained
with sparse matrix diagonalization. 
The difference between the second
order and exact results indicate the contributions from the third
and higher orders, which are more than one order smaller in magnitude.

For $^{16}$O we cannot obtain non-perturbative results for benchmarking
due to the severe sign problem, and so instead we vary the zeroth
order Hamiltonian to triangulate the binding energy and estimate its
uncertainty. In Fig.~\ref{fig:lambda_dependence}(c)
the $\dagger$ and $\ddagger$ symbols mark the ptQMC energies calculated
with $H_{0}=K+1.1V_{0}$ and $H_{0}=K+1.2V_{0}$, respectively. For
each calculation, we use $V_{C}=H-H_{0}$ as the perturbing Hamiltonian and plot the
energies as functions of the small parameter $\lambda$. 
While the variation of $H_{0}$ shifts the zeroth
order energy by about 50~MeV, for full Hamiltonian $H$ ($\lambda=1$)
we find that the first and second order energies
only vary by about 4~MeV and 2.4~MeV, respectively.
These variations can be identified as the truncation errors of the perturbative
series at corresponding orders, see also \cite{SM}.

In Tab.~\ref{tab:The-nuclear-binding} we present the ptQMC energies
for several nuclei compared to the empirical values.
The improvement of
$E_{2}$ compared with $E_{{\rm 1}}$ is clearly seen. Generally,
the correlation energy $\delta E_{2}$ accounts for about $20\%$
of the total binding energy for all nuclei with $A\geq4$. We note
that the first order energy is the expectation value of the full Hamiltonian
using the zeroth order wave function $|\Psi_{0}\rangle$, and it is
an upper bound on the ground state energy. The energy correction
$\delta E_{2}$ is negative definite, reflecting
the fact that the corrected wave function $|\Psi_{1}\rangle$ is
much closer to the exact ground state than $|\Psi_{0}\rangle$.

In perturbative calculations the convergence pattern can be invalidated by symmetry constraints.
As the unperturbed Hamiltonian
$H_{0}$ respects the SU(4) symmetry, the wave function $|\Psi_{0}\rangle$
must belong to one of its irreducible representations (\textit{irreps}).
The full Hamiltonian breaks the SU(4) symmetry, thus its ground state $|\Psi\rangle$ is a
mixture of different SU(4) \textit{irreps}.
As is shown in Fig.~\ref{fig:lambda_dependence}(d),
the components of $|\Psi\rangle$ that mixes the SU(4) \textit{irreps} can only
be seen in $|\delta\Psi_{1}\rangle$ or $\delta E_2$.
This explains the large $\delta E_2$ in $^{16}$O that can not be eliminated by varying $H_0$.
We note that this effect is strongest for the OPEP in Eq.~(\ref{eq:V2N}) as it breaks
both the Wigner-SU(4) and the spin SU(2) symmetries.

\begin{table}
\caption{\label{tab:The-nuclear-binding}Binding energies at different
orders calculated with  ptQMC. compared to experiment (all in MeV).
The errors are combinations of MC statistical errors and extrapolation errors~\cite{SM}.
See Fig.~2(c) for further notations.}
\centering{}%
\begin{tabular}{ccccccc}
\hline
 & $E_{0}$  & $\delta E_{1}$  & $E_{1}$  & $\delta E_{2}$  & $E_{2}$  & $E_{{\rm exp}}$\tabularnewline
\hline
$^{3}$H  & $-7.41(3)$  & $+2.08$  & $-5.33(3)$  & $-2.99$  & $-8.32(3)$  & $-8.48$\tabularnewline
$^{4}$He  & $-23.1(0)$  & $-0.2$  & $-23.3(0)$  & $-5.8$  & $-29.1(1)$  & $-28.3$\tabularnewline
$^{8}$Be  & $-44.9(4)$  & $-1.7$  & $-46.6(4)$  & $-11.1$  & $-57.7(4)$  & $-56.5$\tabularnewline
$^{12}$C  & $-68.3(4)$  & $-1.8$  & $-70.1(4)$  & $-18.8$  & $-88.9(3)$  & $-92.2$\tabularnewline
$^{16}$O  & $-94.1(2)$  & $-5.6$  & $-99.7(2)$  & $-29.7$  & $-129.4(2)$  & $-127.6$\tabularnewline
$^{16}$O$^{\dagger}$  & $-127.6(4)$  & $+24.2$  & $-103.4(4)$  & $-24.3$  & $-127.7(2)$  & $-127.6$\tabularnewline
$^{16}$O$^{\ddagger}$  & $-161.5(1)$  & $+56.8$  & $-104.7(2)$  & $-22.3$  & $-127.0(2)$  & $-127.6$\tabularnewline
\hline
\end{tabular}
\end{table}


In summary, we have presented a novel algorithm (ptQMC) that allows for a
precise calculation of the second order perturbative correction in QMC
without referring to the full spectrum of the
excited states. While the QMC method with simplified interactions are successfully applied
in various fields of physics
~\cite{Wiringa_2002PRL, Lu2019_PLB,Sandvik_2007PRL, Wang_2015PRL, Wei_2016PRL, Li_2016PRL,Wu_2003PRL, Wu_2005PRB, Bulgac_2006PRL, Halford_2020PRL,Umrigar_2007PRL, Hangleiter_2020SA},
attempts to use more
realistic interactions are hindered by the sign problem. The
ptQMC method is free from sign problems and opens the way to
treat complex interactions systematically.
Our method converges quickly for relatively soft interactions.
For interactions with strong short-distance correlations such as tensor forces, which are important in electroweak processes~\cite{Menendez_2011PRL},
some pre-processing of the interaction using  renormalization group transformations or some analogous method is required.

\paragraph{Acknowledgements}

We are grateful for discussions with members of the Nuclear Lattice
Effective Field Theory Collaboration. We gratefully acknowledge funding
by NSAF (Grant No. U1930403), the Deutsche Forschungsgemeinschaft
(DFG, German Research Foundation) and the NSFC through the funds provided
to the Sino-German Collaborative Research Center TRR110 ``Symmetries
and the Emergence of Structure in QCD'' (DFG Project ID 196253076
- TRR 110, NSFC Grant No. 12070131001), the Chinese Academy of Sciences
(CAS) President's International Fellowship Initiative (PIFI) (Grant
No. 2018DM0034), Volkswagen Stiftung (Grant No. 93562), the European
Research Council (ERC) under the European Union's Horizon 2020 research
and innovation programme (grant agreement No. 101018170) and the U.S.
Department of Energy (DE-SC0013365 and DE-SC0021152) and the Nuclear
Computational Low-Energy Initiative (NUCLEI) SciDAC-4 project (DE-SC0018083)
and the Scientific and Technological Research Council of Turkey (TUBITAK
project no. 120F341) and the National Natural Science Foundation of China under Grants No. 12105106 and the China Postdoctoral Science Foundation under Grant No. BX20200136, 2020M682747 as well as computational resources provided by
the Beijing Super Cloud Computing Center (BSCC, http://www.blsc.cn/), TianHe 3F,
the Gauss Centre for Supercomputing e.V. (www.gauss-centre.eu) for
computing time on the GCS Supercomputer JUWELS at J{\"u}lich Supercomputing
Centre (JSC) and the Oak Ridge Leadership Computing Facility through
the INCITE award ``Ab-initio nuclear structure and nuclear reactions''.
Further computational resources from the JSC on JURECA DC are gratefully
acknowledged.

\newpage

\section*{SUPPLEMENTAL MATERIAL}
\addtocounter{section}{10}

In the main text we focus on the perturbative QMC algorithm and its
capability of solving realistic \textit{ab initio} nuclear models.
Here, we provide more details. In Eq.~(11) we claim that
the sign problem in integrating the $c$ field can be alleviated by
shifting the integral contour, here we present a numerical demonstration.
We also give the details of constructing the N$^{2}$LO chiral force.
We further discuss the imaginary time extrapolation.
The deuteron binding energy calculation is used to show that even
though second order corrections can be sizeable, effects from the
third and higher orders can be small (as claimed in the main text).

\subsection{Integral variable change for the $c$ field}

\begin{figure}
\begin{centering}
\includegraphics[width=0.85\columnwidth]{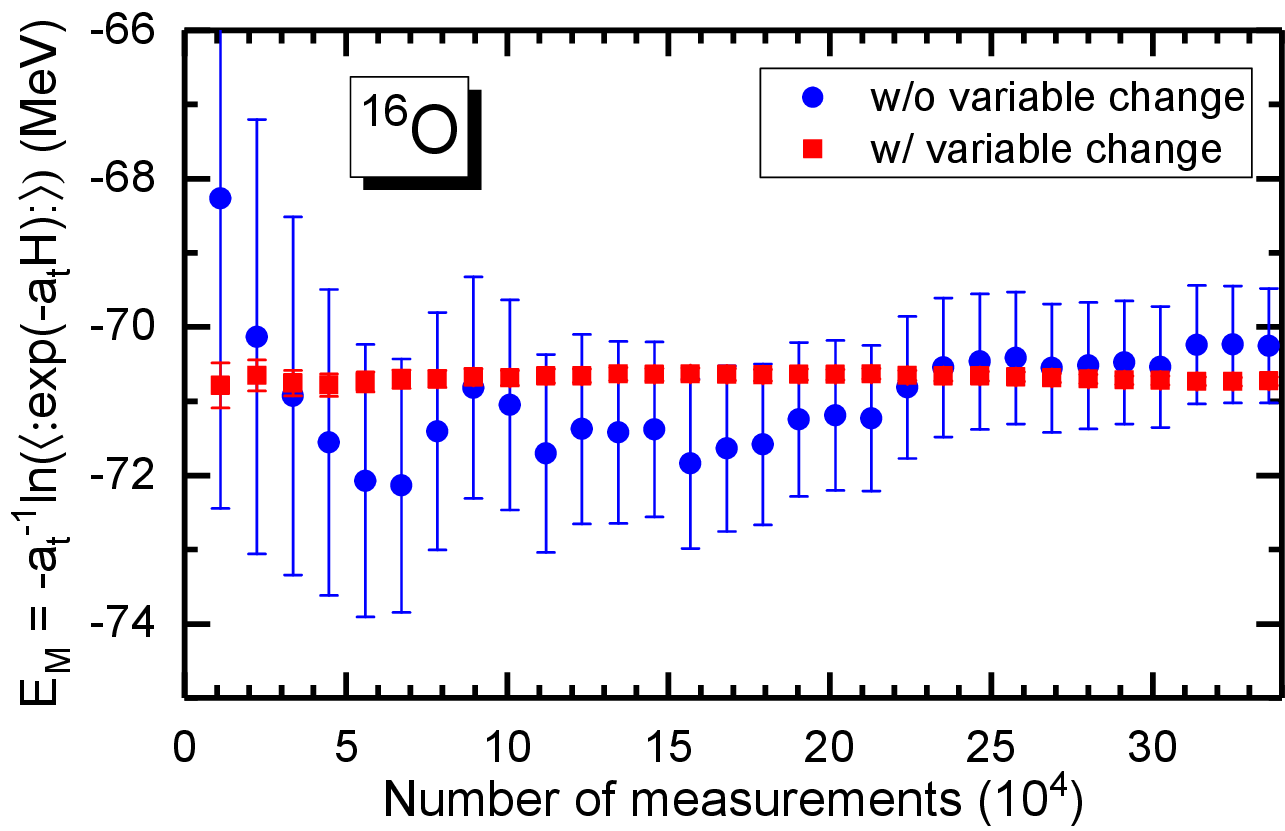}
\par\end{centering}
\begin{centering}
\includegraphics[width=0.85\columnwidth]{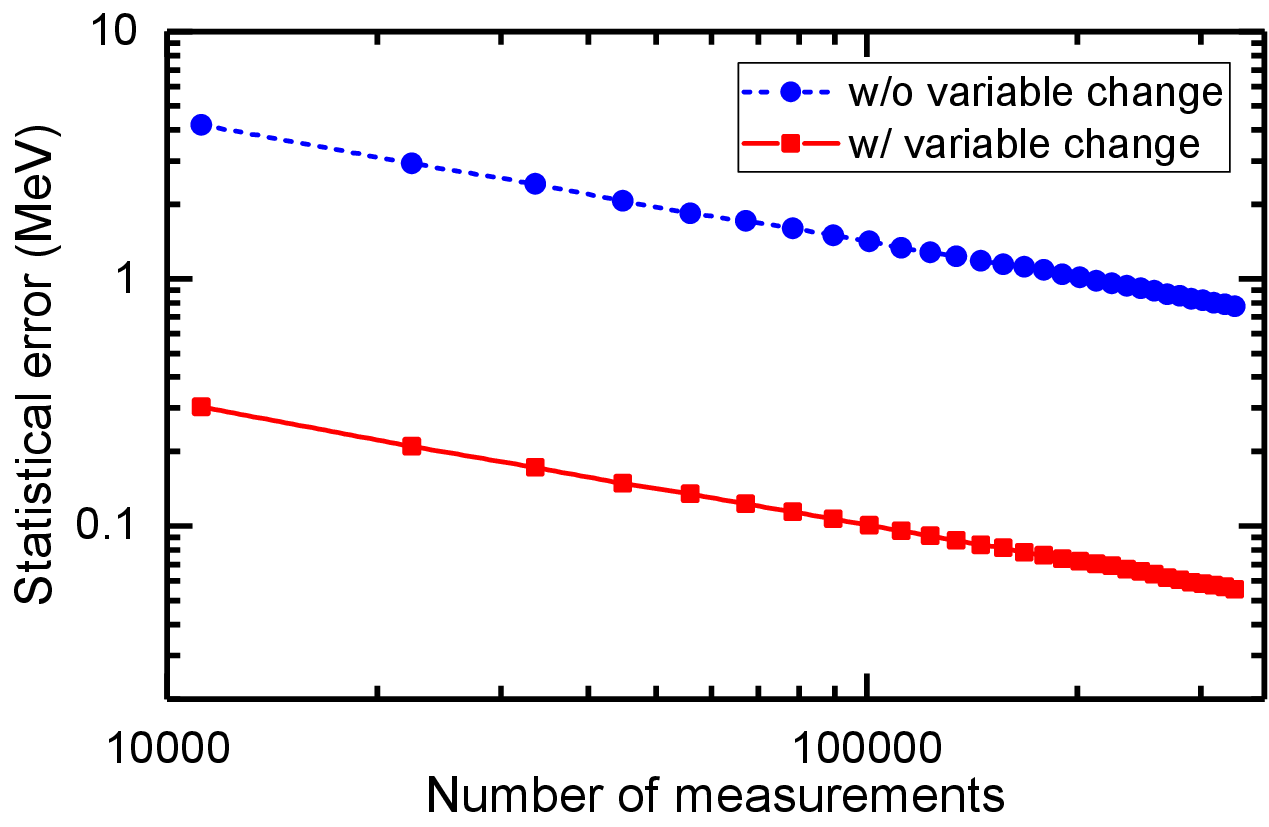}
\par\end{centering}
\caption{\label{fig:benchmark_variable_shift}(Upper panel) Calculated transfer
matrix energies $E_{M}$ as functions of the number of measurements.
The circles (blue) and squares (red) denote the calculations without
and with the variable shift. (Lower panel) The corresponding statistical
errors.}
\end{figure}

In Eq.~(11) we introduced a variable change that can
alleviate the sign problem and reduce the statistical error. Here
we demonstrate this point by comparing the results calculated without
and with the variable change. In the upper panel of Fig.~\ref{fig:benchmark_variable_shift} we
show the calculated transfer matrix energy
\begin{align}
E_{M} & =-\frac{1}{a_{t}}\ln\frac{\langle\Psi_{T}|M|\Psi_{T}\rangle}{\langle\Psi_{T}|\Psi_{T}\rangle}\nonumber \\
 & =-\frac{1}{a_{t}}\ln\frac{\int\mathcal{D}s\mathcal{D}cP(s)P(c)\langle\Psi_{T}|M(s,c)|\Psi_{T}\rangle}{\langle\Psi_{T}|\Psi_{T}\rangle},
\end{align}
where $M$ is the full transfer matrix corresponding to the full N$^{2}$LO
chiral interaction Eqs.(14) in the main text, $s$ and $c$ represent
the auxiliary fields from decomposing the interactions. We take $^{16}$O
as an example and $|\Psi_{T}\rangle$ is  a shell model wave
function (see below). The circles denote the results calculated by sampling the
$s$ and $c$ field directly with a standard normal distribution,
while the squares show results obtained with the variable change.
We see that the latter calculation converges much faster. We also
show the statistical errors in the lower panel. In both cases the
errors decrease according to the theoretical scaling law $\Delta E\propto N^{-1/2}$
with $N$ the number of measurements. With the variable change, however,
the statistical errors are about one order of magnitude smaller.

\subsection{The zeroth order SU(4) Hamiltonian}

\begin{figure}
\begin{centering}
\includegraphics[width=0.4\textwidth]{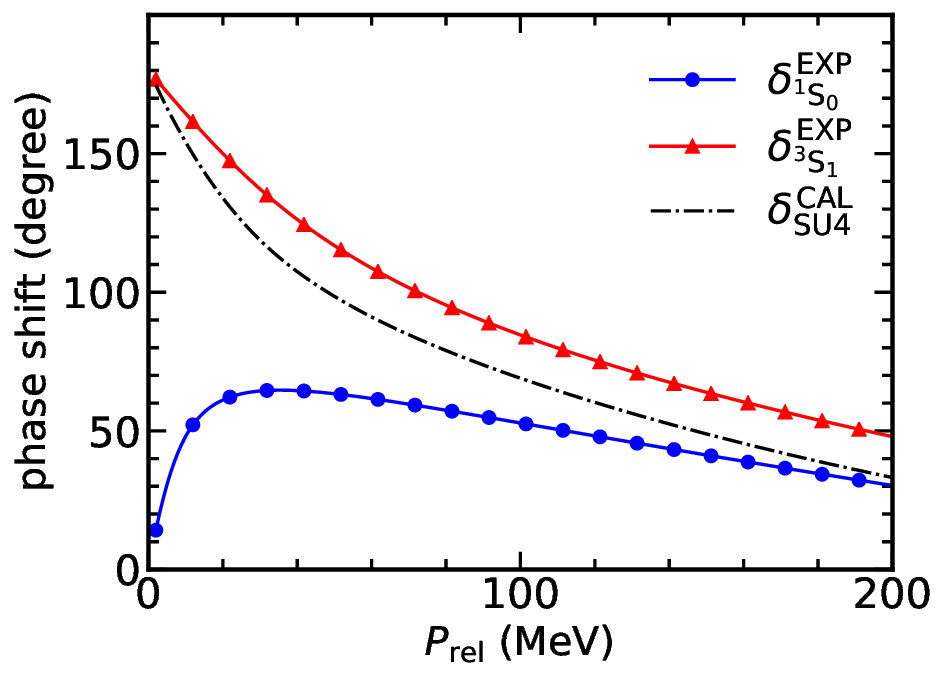}
\par\end{centering}
\caption{\label{fig:zeroth_phase_shifts}The triangles (red) and circles (blue)
denote the empirical $^{1}S_{0}$ and $^{3}S_{1}$ phase shifts, respectively~\cite{Stoks1993}.
The dash-dotted curve represents the results from the zeroth order
Hamiltonian Eq.~(\ref{eq:HSU4}).}
\end{figure}

In the main text we use a zeroth order Hamiltonian that respects the
Wigner-SU(4) symmetry. The details and parametrization can be found
in Ref.\cite{Lu2019_PLB}. For completeness we also present the details
here. On a periodic $L^{3}$ cube with lattice coordinates $\bm{n}=(n_{x,}n_{y},n_{z})$,
the Hamiltonian is
\begin{equation}
H_{0}=K+\frac{1}{2}C_{{\rm SU4}}\sum_{\bm{n}}:\tilde{\rho}^{2}(\bm{n}):,\label{eq:HSU4}
\end{equation}
where $K$ is the kinetic energy term with nucleon mass $m=938.92$
MeV and the $::$ symbol indicate normal ordering. The smeared density
operator $\tilde{\rho}(\bm{n})$ is defined as
\begin{equation}
\tilde{\rho}(\bm{n})=\sum_{i}\tilde{a}_{i}^{\dagger}(\bm{n})\tilde{a}_{i}(\bm{n})+s_{L}\sum_{|\bm{n}^{\prime}-\bm{n}|=1}\sum_{i}\tilde{a}_{i}^{\dagger}(\bm{n}^{\prime})\tilde{a}_{i}(\bm{n}^{\prime}),
\end{equation}
where $i$ is the joint spin-isospin index and the smeared annihilation
and creation operators are defined as
\begin{equation}
\tilde{a}_{i}(\bm{n})=a_{i}(\bm{n})+s_{NL}\sum_{|\bm{n}^{\prime}-\bm{n}|=1}a_{i}(\bm{n}^{\prime})~.
\end{equation}
The summation over the spin and isospin implies that the interaction
is SU(4) invariant. The parameter $s_{L}$ controls the strength of
the local part of the interaction, while $s_{NL}$ controls the strength
of the non-local part of the interaction. Here we include both kinds
of smearing. Both $s_{L}$ and $s_{NL}$ have an impact on the range
of the interactions. The parameter $C_{{\rm SU4}}$ gives the strength
of the two-body interactions. In this work we use a lattice spacing
$a=1.32$~fm and the parameter set $C_{{\rm SU4}}=-3.41\times10^{-7}$~MeV$^{-2}$,
$s_{L}=0.061$ and $s_{NL}=0.5$. These parameters
together with a properly chosen three-body force can reproduce the
binding energies and charge density distributions of light nuclei
from $^{3}$H to the Ca isotopes~ \cite{Lu2019_PLB}. In Fig.~\ref{fig:zeroth_phase_shifts}
we show the NN $S$-wave phase shifts calculated with Eq.~(\ref{eq:HSU4})
(dash-dotted line) compared with the empirical $^{1}S_{0}$ (circles)
and $^{3}S_{1}$ (triangles) phase shifts.

\subsection{Construction of the N$^{2}$LO chiral interaction}

In this section we present the details of the nuclear chiral interaction
used in the main text. Recently we have built a next-to-next-to-next-to-leading-order
(N$^{3}$LO) chiral interaction on the lattice, where all $24$ low-energy
constants (LECs) are determined by fitting to the empirical partial
wave phase shifts and mixing angles \cite{Li2018} based on the improved
spherical wall method \cite{Lu2016}. These interactions are non-local
and difficult to realize efficiently using auxiliary fields. In this
work we employ another set of semi-local contact operators, which
is completely equivalent to the non-local operators in Ref.~\cite{Li2018}
for two-body scattering. We note that similar constructions have already
been used for the Green's function Monte Carlo calculations \cite{Gezerlis2013}.
The local operator basis used here contains isospin dependent terms proportional
to $\bm{\tau}_{1}\cdot\bm{\tau}_{2}$.
The results are given in Eq.~(14).
In this form, these operators can be written as products of one-body
density operators and decomposed using auxiliary field transformations.

For 2N contact terms we introduce an extra non-local regulator $f_{{\rm 2N}}=\exp[-\sum_{i=1}^{2}\left(p_{i}^{6}+p_{i}^{\prime6}\right)/\Lambda^{6}]$,
with $\Lambda=340\,$MeV.
Similarly, the OPEP in the last
line of Eq.(14) is regulated with a local exponential regulator $f_{\pi}(q^{2})=\exp[-(q^{2}+M_{\pi}^{2})/\Lambda_{\pi}^{2}]$
with $\Lambda_{\pi}=300$~MeV.
The cutoffs are chosen to
satisfy $\Lambda,\Lambda_{\pi}\ll\pi/a$ so as to minimize the lattice
artifacts. We also tested other choices of the cutoffs and found similar
results as presented here.
For N$^{2}$LO calculations in this work we fit the
LECs to the Nijmegen phase shifts~\cite{Stoks1993} below $P_{{\rm rel}}$=200 MeV$<$2$M_{\pi}$.
In this momentum interval the two-pion exchange potential can be approximately
absorbed into the contact terms and discarded.

Besides the short-range contact terms, we also need a long-range one-pion-exchange
potential (OPEP). Recently a semi-local momentum space regularized
chiral potential was developed up to fifth order \cite{Reinert2018}.
This regularization method is more convenient than other choices for
lattice simulations. The OPEP we used is given in the last line of
Eq.~(14). The constant $C_{\pi}^{\prime}$ is defined as
\[
C_{\pi}^{\prime}=\frac{1}{3\Lambda_{\pi}^{3}}\Biggl[\Lambda_{\pi}(\Lambda_{\pi}^{2}-2M_{\pi}^{2})+2\sqrt{\pi}M_{\pi}^{3}\exp(\frac{M_{\pi}^{2}}{\Lambda_{\pi}^{2}}){\rm erfc}(\frac{M_{\pi}}{\Lambda_{\pi}})\Biggr].
\]
The term proportional to $C_{\pi}^{\prime}$ is a counterterm introduced
to remove the short-range singularity from the OPEP \cite{Reinert2018}.
We note that the OPEP regulated in this way is soft and adaptive to perturbative calculations.
This can be clearly seen by comparing the contribution of the OPEP $\langle V_{\rm{OPEP}} \rangle$ with the total potential energy $\langle V_{\rm{2N}} + V_{\rm{3N}} + V_{\rm{cou}} \rangle$.
Taking $^{16}$O as an example, in this work we find $\langle V_{\rm{OPEP}} \rangle$ = $-18.9$ MeV, which is more than one order smaller than the total potential energy $-361.1$ MeV.
Thus we expect a fast convergence for the perturbative calculations including the OPEP.
In section F below we will numerically demonstrate this point with the deuteron.

For the three-body force at N$^{2}$LO we adopt a simple 3N contact
term with Wigner SU(4) symmetry,
\begin{equation}
V_{{\rm 3N}}=\frac{c_{E}}{2F_{\pi}^{4}\Lambda_{\chi}}f_{{\rm 3N}}(p_{1},p_{2},\cdots,p_{3}^{\prime}),
\end{equation}
where $\Lambda_{\chi}=700$~MeV is the chiral symmetry breaking scale, $F_\pi=92.2$ MeV is the pion decay constant, $c_E$ is the coupling constant, $f_{{\rm 3N}}=\exp[-\sum_{i=1}^{3}\left(p_{i}^{6}+p_{i}^{\prime6}\right)/\Lambda^{6}]$
is a seperable non-local regulator.
In this work we use the same cutoff $\Lambda=340$ MeV for both 2N and 3N interactions.

Besides the nuclear force we also include a Coulomb force.
With lattice notations we write
\begin{equation}
V_{{\rm cou}} = :\frac{\alpha}{2}\sum_{\bm{n}_{1}\bm{n}_{2}}f_{{\rm c}}(\bm{n}_{1}-\bm{n}_{2})\rho_{p}(\bm{n}_{1})\rho_{p}(\bm{n}_{2}):
\end{equation}
where $\rho_{p}$ is the total proton density. The fine structure
constant $\alpha=1/137$ and the function $f_{{\rm c}}=1/\max(|\bm{n}_{1}-\bm{n}_{2}|,1/2)$
give the regularized Coulomb force.

For a complete calculation, we should also include
the long-range three-body forces from pion-exchange diagrams. However,
these terms have only a minor impact on the main computational analysis
of this study and so is reserved for future work. In $V_{{\rm 2N}}$
and $V_{{\rm 3N}}$, we regulate the single particle momenta instead
of the relative Jacobi momenta. These forms are more convenient to
implement on the lattice but violate Galilean invariance. Nevertheless,
the leading order Galilean breaking effect occurs at $\mathcal{O}((Q/\Lambda)^{6})$
and will not be considered in the N$^{2}$LO calculations presented
here.

\begin{figure*}
\begin{centering}
\includegraphics[width=1\textwidth]{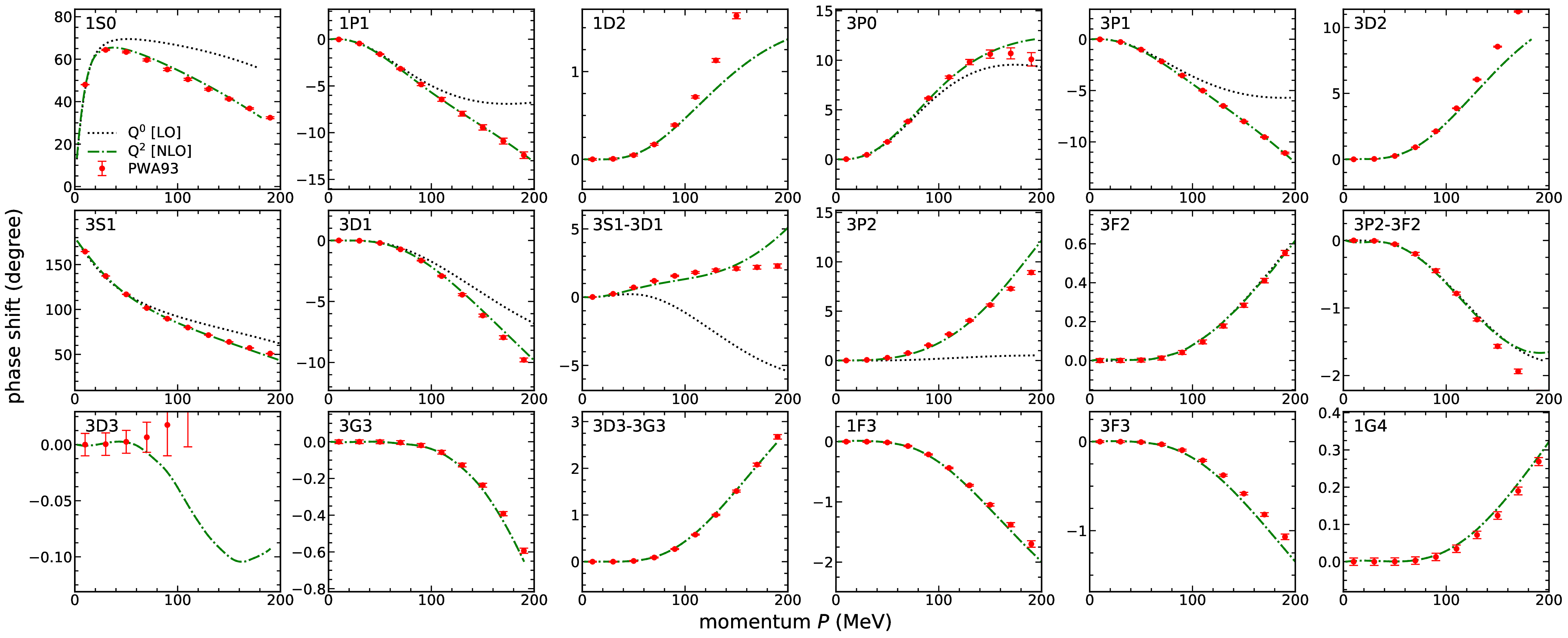}
\par\end{centering}
\caption{Calculated phase shifts for the LECs in Tab.~\ref{tab:Fitted-LECs-at}.
Dotted and dash-dotted lines denote the LO and NLO results, respectively.
Red circles with error bars show the empirical values \cite{Stoks1993}.
\label{fig:Calculated-phase-shifts}}
\end{figure*}

We determine the LECs $B_{i}$, $C_{i}$ and
$c_{E}$ by fitting to the low-energy NN phase shifts, mixing angles
and triton energy. The method is based on Ref.~\cite{Lu2016}. We
decompose the scattering waves on the lattice into different partial
waves, then employ the real and complex auxiliary potentials to extract
the asymptotic radial wave functions. We follow the conventional procedure
for fitting the LECs in the continuum \cite{Epelbaum2005}. We first
determine the spectroscopic LECs for each partial wave, then the $B_{i}$
and $C_{i}$ can be obtained by solving the linear equations.
In Table \ref{tab:Fitted-LECs-at} we show the fitted LECs at
NLO for cutoff $\Lambda=340$ MeV. At this order we consider NN scattering
up to a relative momentum $P_{{\rm rel}}=200$ MeV. Here we use the
lattice unit system $\hbar=c=a=1$ and all LECs are dimensionless.
In Fig.~\ref{fig:Calculated-phase-shifts} we show the calculated
phase shifts. The dotted and dash-dotted lines denote the results
at LO and NLO, respectively. The red dots with error bars are empirical
values from the Nijmegen partial wave analysis (NPWA)\cite{Stoks1993}
.

\begin{table}
\caption{\label{tab:Fitted-LECs-at}Fitted LECs at N$^{2}$LO with $\Lambda=340$~MeV and $\Lambda_\pi=300$~MeV (dimensionless).}
\centering{}

\begin{tabular}{cccccc}
\hline
LEC & $B_{1}$ & $B_{2}$ & $C_{1}$ & $C_{2}$ & $C_{3}$\tabularnewline
\hline
 & $-2.443$ & $-0.125$ & $0.143$ & $-0.012$ & $-0.013$\tabularnewline
\hline
LEC & $C_{4}$ & $C_{5}$ & $C_{6}$ & $C_{7}$ & $c_{E}$\tabularnewline
\hline
 & $-0.020$ & $0.273$ & $0.0$ & $-0.078$ & $0.712$\tabularnewline
\hline
\end{tabular}
\end{table}

\begin{figure}[H]
\begin{centering}
\includegraphics[width=0.4\textwidth]{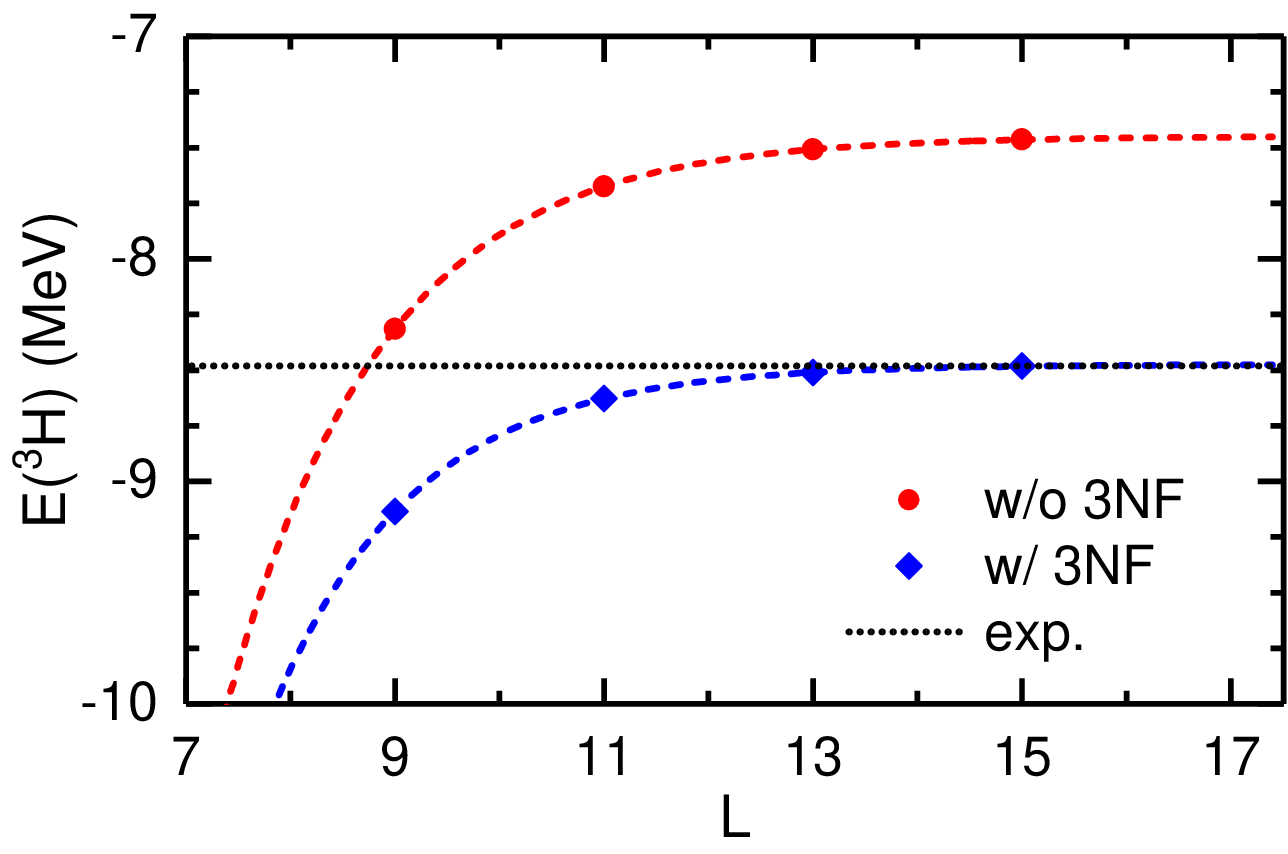}
\par\end{centering}
\caption{\label{fig:Triton-energies-as}Triton binding energies as functions of the
box size $L$. The circles (blue) and diamonds (red) denote the calculations
without and with the three-body force, respectively. Horizontal line denotes the experimental value $E=-8.482$~MeV.}
\end{figure}

We determine the three-body coupling constant $c_{E}$ by
fitting to the triton energy. In Fig.~\ref{fig:Triton-energies-as}
we show the triton energy calculated with the NLO interactions with
the parameters from Table~\ref{tab:Fitted-LECs-at} as blue circles.
We find that the experimental triton energy $E(^{3}$H$)=-8.482$~MeV
can be reproduced with $c_{E}=0.712$ at infinite volume. The corresponding
results are shown as red diamonds. All results for the triton are
obtained by exactly diagonalizing the lattice Hamiltonian using sparse
matrix algebra.

\subsection{Trial wave functions}

In this work the $^{3}$H nucleus was always solved exactly with the
sparse matrix algebra. The other nuclei from $^{4}$He are simulated with
the imaginary time projection method. The projection method requires
properly chosen trial wave function that have large overlap with the
exact ground state. We have tested different choices of the trial
wave functions and found that the clustering states is best suited
for nuclei lighter than $^{16}$O. For $^{16}$O a shell model wave
function works better.

On the lattice the clustering state writes as the anti-symmetrized
product of the single particle wave functions
\[
\psi_{i}(\bm{r},s,t)=\exp\left[-\frac{(\bm{r}-\bm{R}_{i})^{2}}{2c^{2}}\right]\chi_{i}\zeta_{i},
\]
where $c=1.4$ fm, $\chi$ and $\zeta$ are spin and isospin spinors,
respectively. For first four nucleons with different spins and isospins
we take $\bm{R}_{i}=(0,0,0)$ in lattice unit, and for next four nucleons we take
$\bm{R}_{i}=(0,0,1)$, and so on. Then the nucleons form a compact
configurations consisting of $\alpha$-clusters centered around the
origin. In Monte Carlo simulations we randomly move the positions
of these $\alpha$-clusters to form state with zero total angular
momentum and zero total momentum.

The shell model wave function is the anti-symmetrized product of the
harmonic oscillator wave functions
\[
\psi_{i}(\bm{r},s,t)=R_{n_{r}}^{L}(r)\vec{Y}_{JM}^{L}(\Omega)\zeta_{i},
\]
where the right-hand side is the solution of the Schrodinger equation
in a harmonic oscillator with frequency $\hbar\omega=41A^{-1/3}$~MeV.
The spins have been coupled with the spatial angular momenta
to form the spinor wave function $\vec{Y}_{JM}^{L}$. For closed shell
nucleus $^{16}$O the nucleons are injected from the bottom until
the $p$-shell is filled up. The total angular momentum is automatically
coupled to zero. In Monte Carlo simulations we also randomize the
center of the harmonic oscillator to form state with zero total momentum.

Note that the full Hamiltonian is translational and rotational invariant
and the exact ground state has zero total momentum $P$ and angular
momentum $J$. The projections to $P=0$ and $J=0$ increase the overlap
of the trial wave function with the exact ground state and accelerate
the convergence.

\subsection{Imaginary time extrapolation}

In Fig.~1 we performed extrapolations $\tau\rightarrow\infty$
to find the ground state energies. We found that both zeroth and second
order energies can be well fitted by a decaying exponential function
$E(\tau)=E(\infty)+Ce^{-\tau\Delta}$, where $E(\infty)$, $C$ and $\Delta$ are
fitting parameters. We note that $\Delta$ has the physical meaning
of the lowest excitation energy. On the contrary, the first order
energies contain two decaying functions $E(\tau)=E(\infty)+Ce^{-\tau\Delta}+C^{\prime}e^{-\tau\Delta/2}$,
where the last term comes from the expectation value of $V_{C}$ in
$|\Psi_{0}\rangle$.

\begin{figure}[htb]
\begin{centering}
\includegraphics[width=0.9\columnwidth]{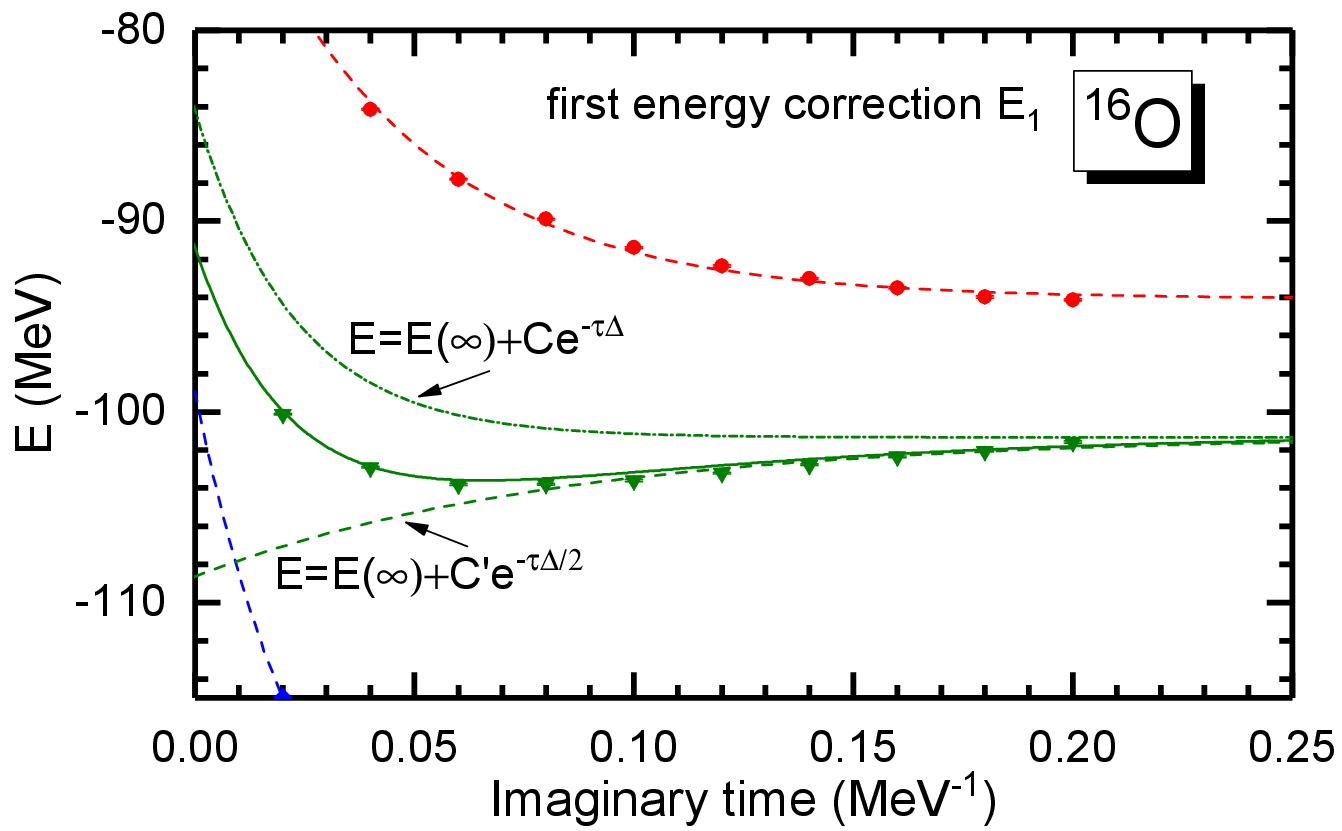}
\par\end{centering}
\caption{\label{fig:first_correction}First energy correction $E_1$ for $^{16}$O
(down triangles). The solid line denotes the fitted result with two decaying exponential functions. The dashed and dot-dashed lines denote the results
with one of the fitting exponential functions removed.}
\end{figure}

In Fig.~1 we observed that the first order energy $E_1$
are not a monotically decreasing function of $\tau$. The reason is
that $E_{1}$ is the expectation value of the full Hamiltonian $H$
in the zeroth order wave function $|\Psi_{0}\rangle$,
\begin{equation}
E_{1}(\tau)=\frac{\langle\Psi_{T}|e^{-\tau H_{0}/2}(H_{0}+V_{C})e^{-\tau H_{0}/2}|\Psi_{T}\rangle}{\langle\Psi_{T}|e^{-\tau H_{0}}|\Psi_{T}\rangle}.\label{eq:first_correction}
\end{equation}
For large $\tau$ only the ground state and first excited state of
$H_{0}$ is relevant. We can approximately write
\begin{align}
e^{-\tau H_{0}/2}|\Psi_{T}\rangle & \rightarrow Ce^{-\tau E_{0}/2}\left[|\Psi_{0}\rangle+C^{\prime}e^{-\tau\Delta/2}|\Psi_{0}^{\prime}\rangle\right]\label{eq:asymptotic}
\end{align}
where $C$ and $C^{\prime}$are certain constants of order $\mathcal{O}(1)$,
the symbols with primes are that for the first excited state of $H_{0}$,
$\Delta$ is the excitation energy. Substituting Eq.~(\ref{eq:asymptotic})
into Eq.~(\ref{eq:first_correction}) and use $\langle\Psi_{0}|\Psi_{0}^{\prime}\rangle=0$,
$\langle\Psi_{0}|\Psi_{0}\rangle=\langle\Psi_{0}^{\prime}|\Psi_{0}^{\prime}\rangle=1$,
we find
\begin{eqnarray}
E_{1}(\tau) & = & \left[E_{0}+e^{-\tau\Delta}|C^{\prime}|^{2}E_{1}+\langle\Psi_{0}|V_{C}|\Psi_{0}\rangle\right.\nonumber \\
 &  & +2e^{-\tau\Delta/2}{\rm Re}\left[C^{\prime}\langle\Psi_{0}^{\prime}|V_{C}|\Psi_{0}\rangle\right]\nonumber \\
 &  & \left.+|C^{\prime}|^{2}e^{-\tau\Delta}\langle\Psi_{0}^{\prime}|V_{C}|\Psi_{0}^{\prime}\rangle\right]/\left(1+|C^{\prime}|^{2}e^{-\tau\Delta}\right)\nonumber \\
 & = & E_{0}+\langle\Psi_{0}|V_{C}|\Psi_{0}\rangle\nonumber \\
 &  & +e^{-\tau\Delta/2}\times2{\rm Re}\left[C^{\prime}\langle\Psi_{0}^{\prime}|V_{C}|\Psi_{0}\rangle\right]\nonumber \\
 &  & +e^{-\tau\Delta}|C^{\prime}|^{2}(E_{1}^{\prime}-E_{1})\label{eq:E1_t_dependence}
\end{eqnarray}
where we omitted the terms decaying faster than $e^{-\tau\Delta}$.
$E_{1}=\langle\Psi_{0}|H|\Psi_{0}\rangle$ and $E_{1}^{\prime}=\langle\Psi_{0}^{\prime}|H|\Psi_{0}^{\prime}\rangle$
are first order energies of the ground state and the excited state,
respectively. The last term in $E_{1}(\tau)$ is positive and produces
the usual exponential decay. However, the term proprotional to $e^{-\tau\Delta/2}$
comes from the matrix element of $V_{C}$ and is not positive definite.
It decays slower and determines the beheaviour of $E_{1}(\tau)$ at
large $\tau$. To see this point more clearly, in Fig.~\ref{fig:first_correction}
we examine the first energy correction $E_{1}$ of $^{16}$O in
more detail. Here we plot the fitting function $E_{1}(\tau)$ as a solid line,
then show the results with one of the two decaying functions removed
seperately. We see that the $e^{-\tau\Delta}$ term decays much faster
and approaches a constant for large $\tau$, while the $e^{-\tau\Delta/2}$
term dominates the asymptotic beheaviour for $\tau>0.1$. As shown
in Eq.~(\ref{eq:E1_t_dependence}), the latter term might be negative
and results in an increasing function at large $\tau$. The summation
of the two exponentials makes the first order energy decrease first
and increasing later, and their cancellation results in a seemingly
fast convergence for $\tau<0.1$. However, this convergence is fictitious
and we need to make the extrapolation more carefully.

On the other hand, the zeroth order energy $E_{0}(\tau)$ is simply
Eq.~(\ref{eq:E1_t_dependence}) with $V_{C}$ set to zero. The term
proportional to $e^{-\tau\Delta/2}$ thus vanishes and we are left with
a simple single decaying exponential function. Further, as we show
in the main text that $E_{2}$ is a good approximation of the exact
energy. Its dependence on $\tau$ is again determined by a single
exponential function. We note that the simple behaviour of $E_{2}(\tau)$
can be seen as a demonstration of the ptQMC method. If the perturbative
corrections are not computed correctly, $E_{2}$ will contain more
complex functions of $\tau$, such as that we see in $E_{1}$.

In fitting the energies we feed the statistical errors from the Monte Carlo
simulations into the Levenberg-Marquardt algorithm.
The resulting uncertainties of the extrapolated energy $E(\infty)$ are adopted
as the errors shown in Fig.~2 and Table I.

\subsection{Convergence of perturbation series beyond second order}\label{sec:F}

In this section we examine the perturbative series beyond second
order using the deuteron as an example. For the deuteron we can solve
the Schr\"odinger equation exactly and find the perturbative corrections
up to very high orders. Here we use the same Hamiltonians $H$ and
$H_{0}$ as used in the main text. By extrapolating to infinite box size $L$,
we find $E(^{2}{\rm H})=-2.28$~MeV for the chiral interaction used in this
work, in good agreement with the experimental value $E_{{\rm exp}}=-2.22$~MeV and within the expected truncation error of the chiral expansion in this order.
As the deuteron binding energy is small, the continuum threshold plays an important role, and the convergence of perturbation theory is not the same as for nuclei with greater binding per nucleon. Thus we will consider a small periodic box $L=5$, for which we have $E(^{2}{\rm H})=-7.733$~MeV. Note that the binding energy per nucleon for medium-mass nuclei
is also of this order.

\begin{figure}[htb]
\begin{centering}
\includegraphics[width=0.8\columnwidth]{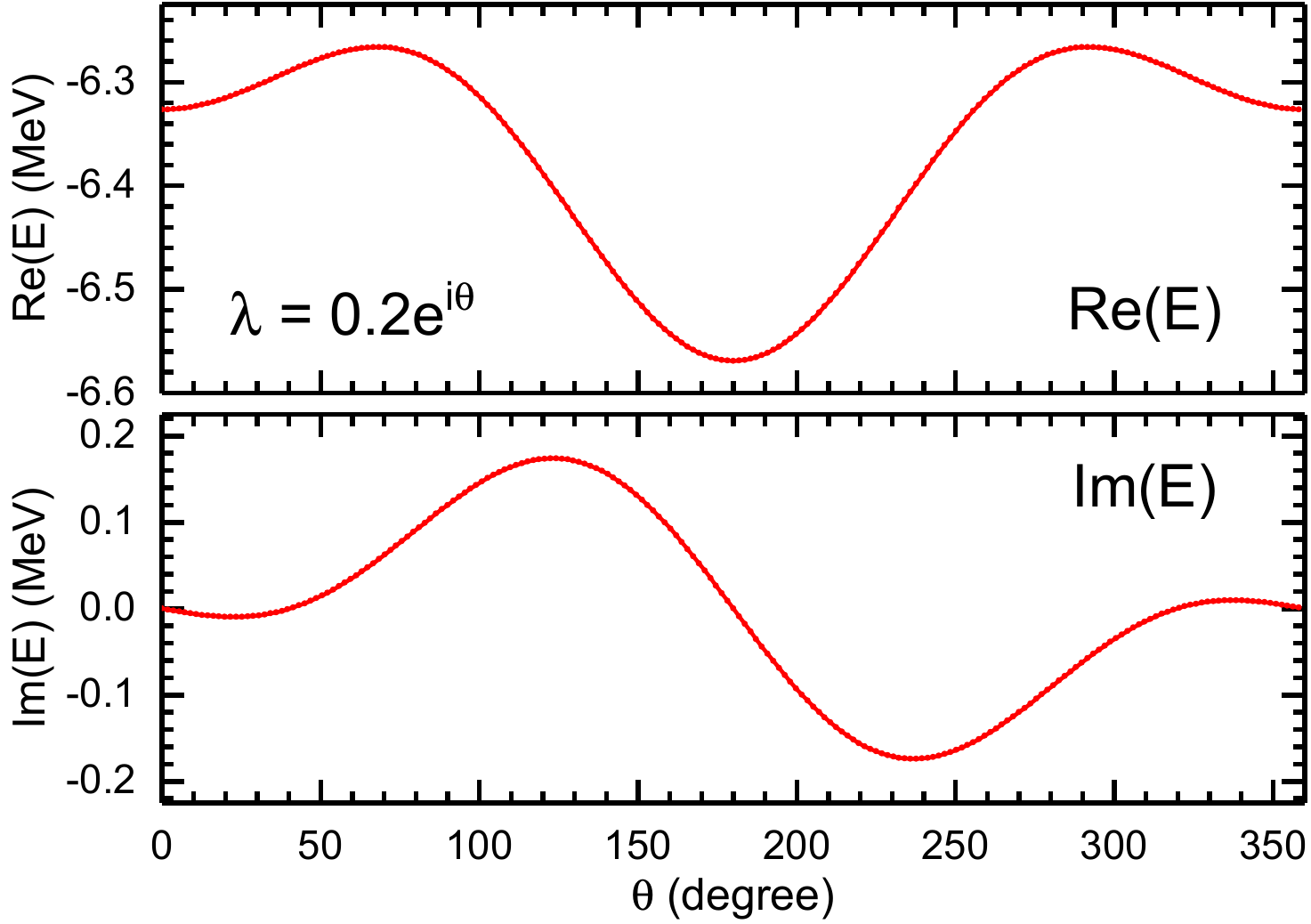}
\par\end{centering}
\caption{\label{fig:Deuteron-energy-as}Deuteron energy as a function of the
parameter $\theta$. Calculated with $\mu=1.0$.}
\end{figure}

We calculate the eigenvalues of the Hamiltonian
\begin{equation}
H=(K+\mu V_{0})+(V_{{\rm 2N}}+V_{{\rm OPEP}}-\mu V_{0}),
\end{equation}
where the symbols are the same as in the main text. Here $\mu$ is
a real constant inserted as an analysis tool. We calculate the term in $K+\mu V_{0}$
using non-perturbative algorithms and treat $V_{{\rm 2N}}+V_{{\rm OPEP}}-\mu V_{0}$
as the perturbing interaction. To obtain the perturbative expansion
precisely, we multiply a variable $\lambda$ to the perturbing Hamiltonian
and calculate the energy $E$ as a function of $\lambda$. We can
use a complex $\lambda$ and calculate $E(\lambda)$ on a closed contour
encircling $\lambda=0$ by exact matrix diagonalization. In Fig.~\ref{fig:Deuteron-energy-as} we show
the real and imaginary parts of $E(\lambda)$ as functions of the
azimuth angle $\theta$ on a circle with the radius $r=0.2$. With
these results we can calculate the derivatives $E^{(n)}(\lambda)$
using Cauchy's formula,
\begin{equation}
E^{(n)}(0)=\frac{n!}{2\pi r^{n}}\int_{0}^{2\pi}E(re^{i\theta})e^{-in\theta}d\theta,\label{eq:Fourier_transform_derivatives}
\end{equation}
which can be performed with discrete Fourier transform. Unlike the
differentiation formulae, the integral formula Eq.~(\ref{eq:Fourier_transform_derivatives})
can be very accurate even for very large $n$. Here we take 200 points uniformly distributed
on the circle and calculate the derivatives up to $n=14$.

\begin{figure}[htb]
\begin{centering}
\includegraphics[width=1\columnwidth]{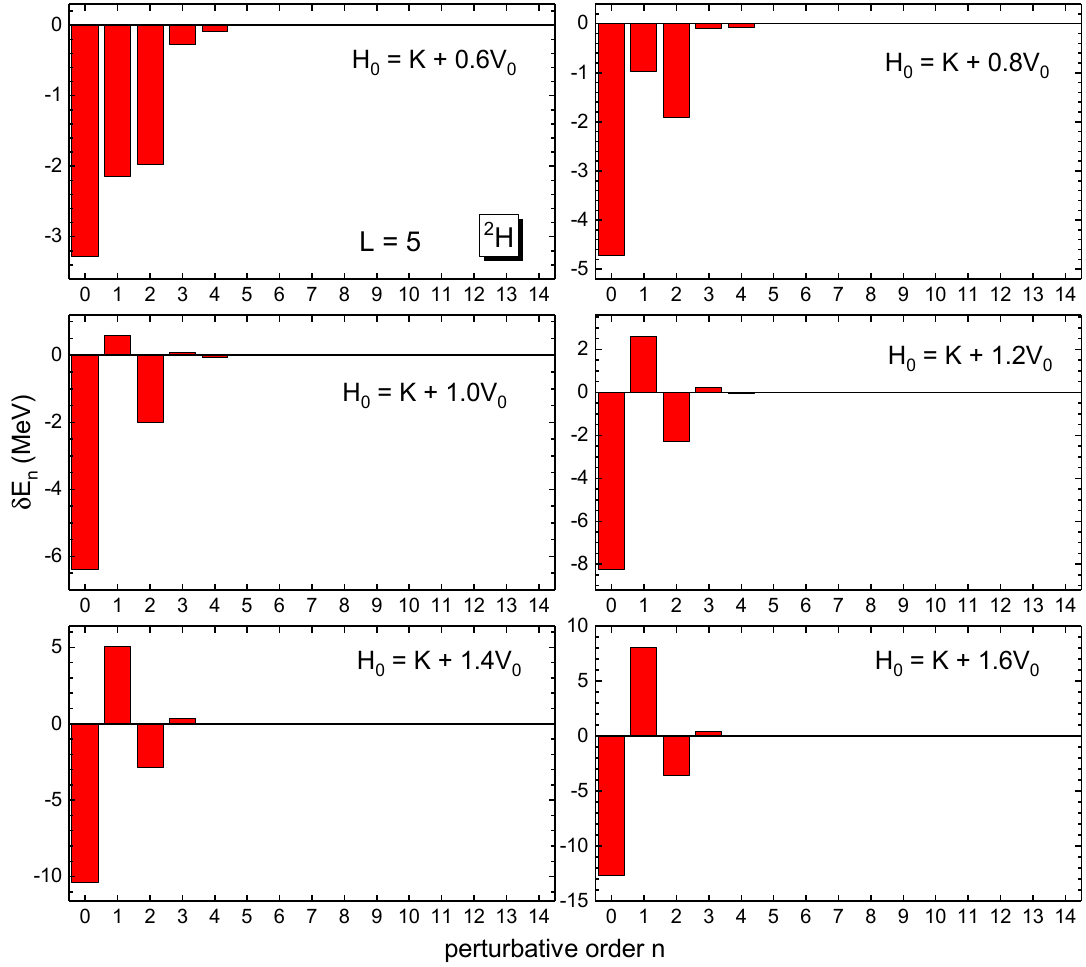}
\par\end{centering}
\caption{\label{fig:Perturbative-energy-corrections}Perturbative energy correction
$\delta E_{n}$ of the deuteron at each order. For the zeroth order we show $E_0$.}
\end{figure}

The energy can be written as a power series,
\begin{equation}
E(\lambda)=\sum_{n=0}^{\infty}\frac{E^{(n)}(0)}{n!}\lambda^{n}.
\end{equation}
Now let us check the convergence pattern of this series. For the full
chiral Hamiltonian we have $\lambda=1$, and the energy correction
at the $n$-th order is simply $\delta E_{n}=E^{(n)}(0)/n!$. In Fig.~\ref{fig:Perturbative-energy-corrections}
we show the energy corrections at each order for six different unperturbed
Hamiltonian corresponding to $\mu=0.6$, $0.8$, $\cdots$, $1.6$,
respectively. We find large $\delta E_{n}$ at the first three orders $n=0$, $1$, $2$.
For $n\geq3$ the contributions are small and become negligible
very quickly when we continue to higher orders. We see that even though
second order correction $\delta E_{2}$ can be large due to symmetry
breaking effects, the third and higher orders follow the normal convergence
pattern.

\begin{figure}[t]
\begin{centering}
\includegraphics[width=1\columnwidth]{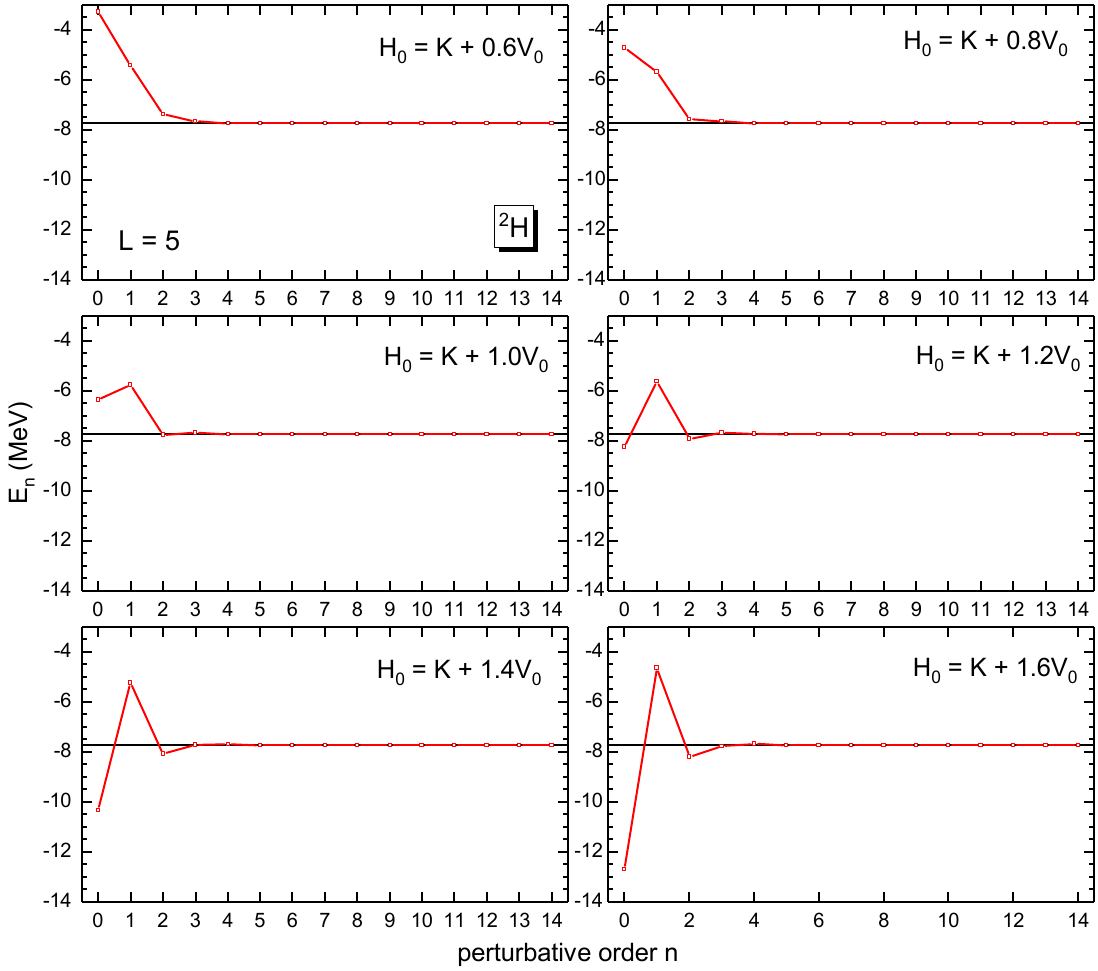}
\par\end{centering}
\caption{\label{fig:Total-energy-of}Partial energy sum $E_{n}$ for the deuteron at each
order. The black horizontal line is the exact energy.}
\end{figure}

Let $E^{(n)}$ be the partial sum of the perturbative energy corrections up to order $n$.  In Fig.~\ref{fig:Total-energy-of} we show
$E_{n}$ versus order $n$ for several different zeroth order Hamiltonians. The
quick convergence to the exact energy can be clearly seen. Though
in some cases the second order energy $E_{2}$ still has a weak dependence
on $\mu$, we find that for $\mu=1.0$ the second order energy $E_{2}=-7.80$
MeV is already very close to the exact value $E=-7.733$ MeV, and the
third order correction is small. As shown in the main text,
the dependence of the perturbative energies on the zeroth order Hamiltonian
can be used as a diagnostic tool for convergence check. In Fig.~\ref{fig:The-zeroth,-first}
we show the total energy $E_{0}$, $E_{1}$ and $E_{2}$ calculated
with different $\mu$. We find that while $E_{0}$ and $E_{1}$ have a strong
dependence on $\mu$, we always find approximately the same second order energy
$E_{2}$ from different zeroth order Hamiltonians and are
close to the exact energy.

For the deuteron in a small periodic box, we conclude that while the
second order correction is sizable due to symmetry-breaking perturbations, higher orders beyond second order are small.  This is consistent with our findings for heavier nuclei in the main text.  Our use of a relatively low momentum cutoff scale is likely playing an important role in keeping the size of the higher order corrections small.

For more general calculations, it is also possible that the higher orders have
alternate signs and cancel with each other to give a small residual term.
This usually occurs when the energy as a function of the small parameter $\lambda$
has a singular point near the origin on the complex plane.
For example, the Taylor series of $f(\lambda)=1/(1+\lambda)$ at $\lambda=0$ has alternate signs near $\lambda=1$ because it has a pole at $\lambda=-1$.
It is well known that for two-body systems a pole or branch point appears when the bound state becomes a continuum state,
thus it would be safe to apply the perturbation theory to deeply bound states as we do in this work.

\begin{figure}[t!]
\begin{centering}
\includegraphics[width=0.9\columnwidth]{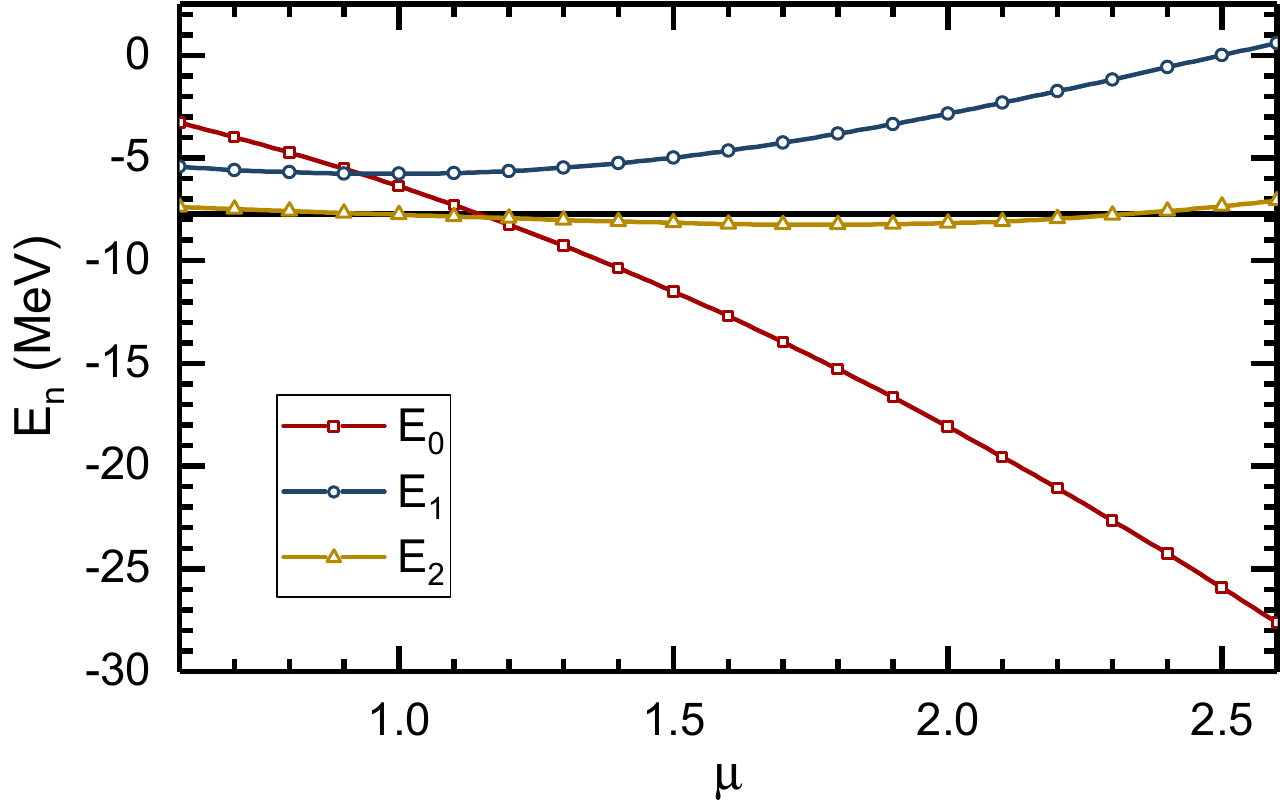}
\par\end{centering}
\caption{\label{fig:The-zeroth,-first}The zeroth, first and second order energies of the deuteron
as functions of the parameter $\mu$. The black horizontal line denotes
the exact energy.}
\end{figure}

\subsection{Truncation errors}
\label{sec:trunc}

In the main text, we have discussed the case of $^{16}$O and triangulated the binding energy.
In this way, we could also obtain an estimate of the truncation error at the given order. Such
an estimate can certainly be improved by referring to the underlying chiral expansion.  A relatively easy way to improve on this uncertainty estimate would be the use of the method proposed in
Ref.~\cite{Epelbaum:2014efa} and refined in Ref.~\cite{LENPIC:2015qsz}. More refined methods are
based on Bayesian or bootstrap methods. For the former type of uncertainty quantification, we
refer to the groundbreaking work in Refs.~\cite{Furnstahl:2015rha,Melendez:2017phj}, as applied e.g.
to neutron-deuteron scattering at higher orders in the chiral EFT in Ref.~\cite{Epelbaum:2019zqc}.


\begin{thebibliography}{10}
\bibitem{Langanke:1995zz} K.~Langanke, D.~J.~Dean, P.~B.~Radha,
Y.~Alhassid and S.~E.~Koonin, Phys. Rev. C \textbf{52}, 718-725
(1995).

\bibitem{Carlson2015_RMP} J. Carlson, S. Gandolfi, F. Pederiva, Steven
C. Pieper, R. Schiavilla, K.~E.~Schmidt, and R.~B.~Wiringa, Rev.
Mod. Phys. 87, 1067 (2015).

\bibitem{Lee2009_PPNP} D. Lee, Prog. Part. Nucl. Phys. 63, 117 (2009).

\bibitem{Lahde:2019npb} T.~A.~L\"ahde and U.-G.~Mei{\ss}ner, Lect.
Notes Phys. \textbf{957}, 1-396 (2019).

\bibitem{Ceperley:1980zz} D.~M.~Ceperley and B.~J.~Alder, Phys.
Rev. Lett. \textbf{45}, 566-569 (1980).

\bibitem{Foulkes2001_RMP} W. M. C. Foulkes, L. Mitas, R. J. Needs,
and G. Rajagopal, Rev. Mod. Phys. 73, 33 (2001).

\bibitem{Assaad:2013xua} F.~F.~Assaad and I.~F.~Herbut, Phys.
Rev. X \textbf{3}, 031010 (2013).

\bibitem{Bulgac:2008zz} A.~Bulgac, J.~E.~Drut and P.~Magierski,
Phys. Rev. A \textbf{78}, 023625 (2008) doi:10.1103/PhysRevA.78.023625.

\bibitem{Carlson:2011kv} J.~Carlson, S.~Gandolfi, K.~E.~Schmidt
and S.~Zhang, Phys. Rev. A \textbf{84}, 061602 (2011).

\bibitem{He:2019ipt} R.~He, N.~Li, B.~N.~Lu and D.~Lee, Phys.
Rev. A \textbf{101}, no.6, 063615 (2020).

\bibitem{Hammond1994} B. J. Hammond, W. A. Lester, and P. J. Reynolds
(1994), Monte Carlo Methods in Ab Initio Quantum Chemistry (World
Scientific, Singapore).

\bibitem{Nightingale1999}M. Nightingale and C. Umrigar (1999), Quantum
Monte Carlo Methods in Physics and Chemistry (Springer). NNDC, (2014),
``Nudat 2,'' http://www.nndc.bnl.gov/nudat2/chartNuc.jsp.

\bibitem{Troyer2005_PRL}M. Troyer, U.-J. Wiese, Phys. Rev. Lett 94,
170201 (2005).

\bibitem{Muroya2003_PTP}S. Muroya, A. Nakamura, C. Nonaka, T. Takaishi,
Prog. Theo. Phys. 110, 615 (2003).

\bibitem{Varney2009_PRB}C. N. Varney, C.-R. Lee, Z. J. Bai, S. Chiesa,
M. Jarrell, and R. T. Scalettar, Phys. Rev. B 80, 075116 (2009).

\bibitem{Wigner1937_PR} E. Wigner, Phys. Rev. 51, 106 (1937).

\bibitem{Elhatisari2016_PRL}S. Elhatisari, N. Li, A. Rokash, J. M.
Alarcon, D. Du, N. Klein, B.-N. Lu, U.-G. Mei{\ss}ner, E. Epelbaum,
H. Krebs, T. A. L\"ahde, D. Lee and G. Rupak, Phys. Rev. Lett. 117,
132501 (2016).

\bibitem{Lu2019_PLB}B.-N. Lu, N. Li, S. Elhatisari, D. Lee, E. Epelbaum,
U.-G. Mei{\ss}ner, Phys. Lett. B 797, 134863 (2019).

\bibitem{Lee2021_PRL} D. Lee, S. Bogner, B. A. Brown, S. Elhatisari,
E. Epelbaum, H. Hergert, M. Hjorth-Jensen, H. Krebs, N. Li, B.-N.
Lu, and U.-G. Mei{\ss}ner, Phys. Rev. Lett. 127, 062501 (2021).

\bibitem{Schwerdtfeger2011_CPC}P. Schwerdtfeger, Chem. Phys. Chem.
12, 3143 (2011).

\bibitem{Epelbaum2009_RMP}E. Epelbaum, H.-W. Hammer, and U.-G. Mei{\ss}ner,
Rev. Mod. Phys. 81, 1773 (2009).

\bibitem{Elhatisari2017_PRL}S. Elhatisari, E. Epelbaum, H. Krebs,
T. A. L{\"a}hde, D. Lee, N. Li, B.-N. Lu, U.-G. Mei{\ss}ner, G. Rupak,
Phys. Rev. Lett. 119, 222505 (2017).

\bibitem{Summerfield2021_PRC}N. Summerfield, B.-N. Lu, C. Plumberg,
D. Lee, J. Noronha-Hostler, A. Timmins, Phys. Rev. C 104, 041901 (2021).

\bibitem{Borasoy2007_EPJA}B. Borasoy, E. Epelbaum, H. Krebs, D. Lee,
U.-G. Mei{\ss}ner, Eur. Phys. J. A 31, 105 (2007).

\bibitem{Borasoy2008_EPJA}B. Borasoy, E. Epelbaum, H. Krebs, D. Lee,
U.-G. Mei{\ss}ner, Eur. Phys. J. A 35, 343 (2008).

\bibitem{Epelbaum2009_EPJA}E. Epelbaum, H. Krebs, D. Lee, U.-G. Mei{\ss}ner,
Eur. Phys. J. A 41, 125 (2009).

\bibitem{Epelbaum2010_PRL}E. Epelbaum, H. Krebs, D. Lee, U.-G. Mei{\ss}ner,
Phys. Rev. Lett. 104, 142501 (2010).

\bibitem{Epelbaum2010_EPJA}E. Epelbaum, H. Krebs, D. Lee, U.-G. Mei{\ss}ner,
Eur. Phys. J. A 45, 335 (2010).

\bibitem{Lahde2014_PLB}T. A. L{\"a}hde, E. Epelbaum, H. Krebs, D.
Lee, U.-G. Mei{\ss}ner, G. Rupak, Phys. Lett. B 732, 110 (2014).

\bibitem{Epelbaum2011_PRL}E. Epelbaum, H. Krebs, D. Lee, U.-G. Mei{\ss}ner,
Phys. Rev. Lett. 106, 192501 (2011).

\bibitem{Epelbaum2012_PRL109}E. Epelbaum, H. Krebs, T. L\"ahde,
D. Lee, U.-G. Mei{\ss}ner, Phys. Rev. Lett. 109, 252501 (2012).

\bibitem{Epelbaum2013_PRL}E. Epelbaum, H. Krebs, T. A. L{\"a}hde,
D. Lee, U.-G. Mei{\ss}ner, Phys. Rev. Lett. 110, 112502 (2013).

\bibitem{Epelbaum2014_PRL}E. Epelbaum, H. Krebs, T. A. L{\"a}hde,
D. Lee, U.-G. Mei{\ss}ner, G. Rupak, Phys. Rev. Lett. 112, 102501
(2014).



\bibitem{Elhatisari2015_Nature}S. Elhatisari, D. Lee, G. Rupak, E.
Epelbaum, H. Krebs, T. A. L{\"a}hde, T. Luu, U.-G. Mei{\ss}ner, Nature
528, 111 (2015).

\bibitem{Li2016_PRL}Z. X. Li, Y. F. Jiang and H. Yao, Phys. Rev.
Lett. 117, 267002 (2016).

\bibitem{Zhang2003_PRL}S. Zhang and H. Krakauer, Phys. Rev. Lett.
90, 136401.

\bibitem{SM} See the {\em Supplemental Material}, which contains details on 
the interaction, the trial wave functions, the imaginary time extrapolation,
details on the perturbative expansion beyond second order and methods
to better estimate the theoretical uncertainties;

\bibitem{Reinert2018_EPJA}P. Reinert, H. Krebs, E. Epelbaum, Eur.
Phys. J. A 54, 86 (2018).

\bibitem{Koenig2017_PRL}S. K{\"o}nig, H. W. Grie{\ss}hammer, H.-W.
Hammer, and U. van Kolck, Phys. Rev. Lett. 118, 202501 (2017).

\bibitem{Koenig2020_EPJA}S. K{\"o}nig, Eur. Phys. J. A 56, 113 (2020).

\bibitem{Vanasse2017_FBS}J. Vanasse and D. R. Phillips, Few-Body
Syst. 58, 26 (2017).

\bibitem{Tjon1975_PLB}J. A. Tjon, Phys. Lett. B 56, 217 (1975).

\bibitem{Platter:2004zs} L.~Platter, H.~W.~Hammer and U.-G.~Mei{\ss}ner,
Phys. Lett. B 607, 254 (2005).


\bibitem{Wiringa_2002PRL} R. B. Wiringa and S. C. Pieper, Phys. Rev. Lett. 89, 182501 (2002).

\bibitem{Sandvik_2007PRL} A. W. Sandvik, Phys. Rev. Lett. 98, 227202 (2007).

\bibitem{Wang_2015PRL} L. Wang, Y. H. Liu, M. Iazzi, M. Troyer and G. Harcos, Phys. Rev. Lett. 115, 250601 (2015).

\bibitem{Wei_2016PRL} Z. C. Wei, C. J. Wu, Y. Li, S. W. Zhang and T. Xiang, Phys. Rev. Lett. 116, 250601 (2016).

\bibitem{Li_2016PRL} Z. X. Li, Y. F. Jiang and H. Yao, Phys. Rev. Lett. 117, 267002 (2016).


\bibitem{Wu_2003PRL} C. J. Wu, J. P. Hu and S. C. Zhang, Phys. Rev. Lett. 91, 186402 (2003).

\bibitem{Wu_2005PRB} C. J. Wu and S. C. Zhang, Phys. Rev. B 71, 155115 (2005).

\bibitem{Bulgac_2006PRL} A. Bulgac, J. E. Drut and P. Magierski, Phys. Rev. Lett. 96, 090404 (2006).

\bibitem{Halford_2020PRL} A. Richie-Halford, J. E. Drut and A. Bulgac, Phys. Rev. Lett. 125, 060403 (2020).

\bibitem{Umrigar_2007PRL} C. J. Umrigar, J. Toulouse, C. Filippi, S. Sorella and R. G. Hennig, Phys. Rev. Lett. 98, 110201 (2007).

\bibitem{Hangleiter_2020SA} D. Hangleiter, I. Roth, D. Nagaj and J. Eisert, Sci. Adv. 6, eabb8341 (2020).

\bibitem{Menendez_2011PRL} J. Men\'{e}ndez, D. Gazit and A. Schwenk, Phys. Rev. Lett. 107, 062501 (2011)







\end{thebibliography}

\begin{thebibliography}{1}
\bibitem{Lu2019_PLB}B.-N. Lu, N. Li, S. Elhatisari, D. Lee, E. Epelbaum,
U.-G. Mei{\ss}ner, Phys. Lett. B 797, 134863 (2019)

\bibitem{Stoks1993}V. G. J. Stoks, R. A. M. Klomp, M. C. M. Rentmeester,
and J. J. de Swart, Phys. Rev. C48, 792 (1993).

\bibitem{Li2018}N. Li, S. Elhatisari, E. Epelbaum, D. Lee, B.-N.
Lu, U.-G. Mei{\ss}ner, Phys. Rev. C 98, 044002 (2018)

\bibitem{Lu2016}B.-N. Lu, T. A. L{\"a}hde, D. Lee, U.-G. Mei{\ss}ner,
Phys. Lett. B 760, 309 (2016)

\bibitem{Gezerlis2013}A. Gezerlis, I. Tews, E. Epelbaum, S. Gandolfi,
K. Hebeler, A. Nogga, and A. Schwenk, Phys. Rev. Lett. 111, 032501
(2013)

\bibitem{Reinert2018}P. Reinert, H. Krebs, E. Epelbaum, Eur. Phys.
J. A 54, 86 (2018)

\bibitem{Epelbaum2005}E. Epelbaum, W. Gl{\"o}ckle, U.-G. Mei{\ss}ner,
Nucl. Phys. A 47, 362 (2005)

\bibitem{Epelbaum:2014efa}
E.~Epelbaum, H.~Krebs and U.-G.~Mei\ss{}ner,
Eur. Phys. J. A \textbf{51}, no.5, 53 (2015).

\bibitem{LENPIC:2015qsz}
S.~Binder \textit{et al.} [LENPIC],
Phys. Rev. C \textbf{93}, no.4, 044002 (2016).

\bibitem{Furnstahl:2015rha}
R.~J.~Furnstahl, N.~Klco, D.~R.~Phillips and S.~Wesolowski,
Phys. Rev. C \textbf{92}, no.2, 024005 (2015).

\bibitem{Melendez:2017phj}
J.~A.~Melendez, S.~Wesolowski and R.~J.~Furnstahl,
Phys. Rev. C \textbf{96}, no.2, 024003 (2017).

\bibitem{Epelbaum:2019zqc}
E.~Epelbaum, J.~Golak, K.~Hebeler, H.~Kamada, H.~Krebs, U.-G.~Mei\ss{}ner, A.~Nogga, P.~Reinert, R.~Skibi\'nski and K.~Topolnicki, \textit{et al.}
Eur. Phys. J. A \textbf{56}, no.3, 92 (2020).

\end{thebibliography}
\end{document}